\documentclass[12pt,preprint]{aastex}
\usepackage{psfig,lscape} 

\def\hal{H$\alpha$}
\def\be{\begin{equation}}
\def\ee{\end{equation}}

\def\m{~$\mu$m}

\def\NII{[\ion{N}{2}]}

\def\ISO{{\it ISO}}
\def\IRAS{{\it IRAS}}
\def\Spitzer{{\it Spitzer}}
\def\2MASS{{\it 2MASS}}
\def\SCUBA{{\it SCUBA}}

\def\Spitzercolora {$f_\nu (70 \mu {\rm m})   / f_\nu (160 \mu {\rm m})$}
\def\Spitzercolorb {$f_\nu (24 \mu {\rm m})   / f_\nu (70 \mu {\rm m})$}
\def\Spitzercolorc {$f_\nu (8.0 \mu {\rm m})  / f_\nu (24 \mu {\rm m})$}

\def\Spitzercolord {$f_\nu (3.6 \mu {\rm m})  / f_\nu (8.0 \mu {\rm m})$}
\def\Spitzercolore {$f_\nu (24 \mu {\rm m})   / f_\nu (160 \mu {\rm m})$}
\def\Spitzercolorf {$f_\nu (70 \mu {\rm m})   / f_\nu (24 \mu {\rm m})$}

\def\N_MIPSIRAC{71}

\begin {document}

\title{Infrared Spectral Energy Distributions of Nearby Galaxies}
  
\author{D.A.~Dale\altaffilmark{1}, G.J.~Bendo\altaffilmark{2}, C.W.~Engelbracht\altaffilmark{2}, K.D.~Gordon\altaffilmark{2}, M.W.~Regan\altaffilmark{3}, L.~Armus\altaffilmark{4}, J.M.~Cannon\altaffilmark{5}, D.~Calzetti\altaffilmark{3}, B.T.~Draine\altaffilmark{6}, G.~Helou\altaffilmark{4}, R.D.~Joseph\altaffilmark{7}, R.C.~Kennicutt\altaffilmark{2}, A.~Li\altaffilmark{8}, E.J.~Murphy\altaffilmark{9}, H.~Roussel\altaffilmark{4}, F.~Walter\altaffilmark{5}, H.M.~Hanson\altaffilmark{1}, D.J.~Hollenbach\altaffilmark{10}, T.H.~Jarrett\altaffilmark{4}, L.J.~Kewley\altaffilmark{7}, C.A.~Lamanna\altaffilmark{1}, C.~Leitherer\altaffilmark{3}, 
M.J.~Meyer\altaffilmark{3}, G.H.~Rieke\altaffilmark{2}, M.J.~Rieke\altaffilmark{2}, K.~Sheth\altaffilmark{4}, J.D.T.~Smith\altaffilmark{2}, M.D.~Thornley\altaffilmark{11}}
\altaffiltext{1}{\scriptsize Department of Physics and Astronomy, University of Wyoming, Laramie, WY 82071; ddale@uwyo.edu}
\altaffiltext{2}{\scriptsize Steward Observatory, University of Arizona, 933 North Cherry Avenue, Tucson, AZ 85721}
\altaffiltext{3}{\scriptsize Space Telescope Science Institute, 3700 San Martin Drive, Baltimore, MD 21218}
\altaffiltext{4}{\scriptsize California Institute of Technology, MC 314-6, Pasadena, CA 91101}
\altaffiltext{5}{\scriptsize Max Planck Institut f\"{u}r Astronomie, K\"{o}nigstuhl 17, 69117 Heidelberg, Germany}
\altaffiltext{6}{\scriptsize Princeton University Observatory, Peyton Hall, Princeton, NJ 08544}
\altaffiltext{7}{\scriptsize Institute for Astronomy, 2680 Woodlawn Drive, Honolulu, HI 96822}
\altaffiltext{8}{\scriptsize Department of Physics and Astronomy, University of Missouri, Columbia, MO 65211}
\altaffiltext{9}{\scriptsize Department of Astronomy, Yale University, New Haven, CT 06520}
\altaffiltext{10}{\scriptsize NASA/Ames Research Center, MS 245-6, Moffett Field, CA 94035}
\altaffiltext{11}{\scriptsize Department of Physics, Bucknell University, Lewisburg, PA 17837}

\begin {abstract}
The \Spitzer\ Infrared Nearby Galaxies Survey (SINGS) is carrying out a comprehensive multi-wavelength survey on a sample of 75 nearby galaxies.  The 1-850~\micron\ spectral energy distributions are presented 
using broadband imaging data from \Spitzer, \2MASS, \ISO, \IRAS, and \SCUBA.  The infrared colors derived from the globally-integrated \Spitzer\ data are generally consistent with the previous generation of models that were developed based on global data for normal star-forming galaxies, though significant deviations are observed.  
\Spitzer's excellent sensitivity and resolution also allow a detailed investigation of the infrared spectral energy distributions for various locations within the three large, nearby galaxies NGC~3031 (M~81), NGC~5194 (M~51), and NGC~7331.  A wide variety of spectral shapes is found within each galaxy, especially for NGC~3031, the closest of the three targets and thus the galaxy for which the smallest spatial scales can be explored.  Strong correlations exist between the local star formation rate and the infrared colors \Spitzercolora\ and \Spitzercolore, suggesting that the 24 and 70\m\ emission are useful tracers of the local star formation activity level.  
Preliminary evidence indicates that variations in the 24\m\ emission, and not variations in the emission from polycyclic aromatic hydrocarbons at 8\m, drive the variations in the \Spitzercolorc\ colors within NGC~3031, NGC~5194, and NGC~7331.  If the galaxy-to-galaxy variations in spectral energy distributions seen in our sample are representative of the range present at high redshift then extrapolations of total infrared luminosities and star formation rates from the observed 24\m\ flux will be uncertain at the factor-of-five level (total range).  The corresponding uncertainties using the redshifted 8.0\m\ flux (e.g. observed 24\m\ flux for a $z = 2$ source) are factors of 10--20.  Considerable caution should be used when interpreting such extrapolated infrared luminosities.
\end {abstract}
 
 
\section {Introduction}

Emission from interstellar dust comprises a wide-ranging portion of the overall energy budget in galaxies, ranging from typically a few percent for ellipticals (Xilouris et al. 2004), to tens of percent for normal star-forming galaxies (Dale et al. 2000), to greater than 95 percent for many Ultraluminous Infrared Galaxies (Soifer et al. 1984).  To take advantage of this energetically-important view of galaxies, many current and planned instruments are pushing infrared, submillimeter, and millimeter astronomy into new sensitivity regimes.  Increasingly deep cosmological surveys carried out at these long wavelengths 
are already revealing a rich diversity of spectral energy distributions (SEDs) at every redshift (e.g. Yan et al. 2004; Rowan-Robinson et al. 2005).  In order to exploit these surveys to their full potential, 
it is paramount to empirically characterize the details of individual galaxy infrared SEDs and their broad characteristics for a diverse sample of galaxies.  

The 75 galaxies studied in detail in the SINGS Legacy Science Program represent a fair sampling of the range of galaxies found in the Local Universe (Kennicutt et al. 2003).  The suite of {\it Spitzer Space Telescope} cosmology surveys, on the other hand, will produce vast catalogs of infrared-bright sources at much larger distances.  Tremendous advances in the understanding of galaxy infrared SEDs could potentially be made with such huge catalogs.  However, due to sensitivity and confusion limitations, these surveys will not constrain the far-infrared 
fluxes for most sources (the initial SWIRE catalog has detections at 70 and 160\m\ for only 0.7\% of 
its sources with photometric redshifts; Rowan-Robinson et al. 2005),
whereas most SINGS targets will have robust detections at near-, mid-, and far-infrared wavelengths in addition to submillimeter data for a third of the sample.  In other words, though it only comprises 75 targets, the broad multi-wavelength coverage for the SINGS sample should enable unique advances in SED understanding.  In addition, inferring a galaxy's total infrared luminosity from just 24\m\ or 24+70\m\ measurements alone would be useful to cosmology surveys, and the collection of SINGS SEDs make it possible to calibrate these extrapolations and quantify the uncertainties involved.  Finally, another important advantage to the SINGS sample lies in its proximity.  The unprecedented combination of \Spitzer\ sensitivity and angular resolution, coupled with the nearness of the SINGS targets, will enable us to learn about differences in the infrared SEDs as a function of the local (sub-kpc to kpc) environment.  The various cosmological efforts being pursued via the {\it Spitzer Space Telescope} will profit from this compilation of SINGS infrared spectra.  

The focus of this study is to make a preliminary exploration of the broad variations of SINGS 1-850\m\ SEDs using both globally-integrated data and data stemming from smaller spatial scales within galaxies.
A complementary study of SINGS infrared spectra will concentrate on a narrower wavelength range to quantify the variations observed in the mid-infrared spectral shapes.


\section {The Sample}

The full SINGS sample is described in Kennicutt et al. (2003).  Mid- and far-infrared images are now becoming available for a good portion of the SINGS sample.  The galaxies chosen for study in this work are simply those first SINGS galaxies with \Spitzer\ mid- and far-infrared broadband data in hand. 
The 
 sample spans a wide range in the infrared/optical ratio and morphological type.  In this work we also present results on the infrared SEDs for various locations within three SINGS galaxies: NGC~3031 (M~81), NGC~5194 (M~51), and NGC~7331.  These three galaxies span large enough solid angles to allow many local environments to be explored.  Figures~\ref{fig:apertures1} and \ref{fig:apertures2} indicate the particular regions selected for analysis in each galaxy overlaid on (unsmoothed) 8.0\m\ images.  Apertures were selected using 8.0 and 24\m\ images.  The diameter of each aperture is at least as large as 38\arcsec, the Full Width at Half Maximum (FWHM) of the poorest resolution Point Spread Function in the dataset (i.e. the 160\m\ beam).  Each image was smoothed to the 160\m\ resolution before fluxes were extracted (\S~\ref{sec:local}).  ``Nuclear'' apertures are centered on the brightest central location; 
``arm'' regions follow the bright regions of the spiral patterns; ``inter-arm'' regions sample the diffuse emission between spiral arms; ``inner disk'' and ``outer disk'' apertures are centered on areas that appear to be a combination of arm and inter-arm regions.

\section {Observations and Data Reduction}
\label{sec:data}
The broad scope of the SINGS imaging program is described in Kennicutt et al. (2003), though a few modifications have since been incorporated, based on our initial validation data for NGC~7331 (e.g. Regan et al 2004).  In brief, mid-infrared mosaics centered on each source are executed for all four IRAC channels (Fazio et al. 2004).  The mosaics are large enough and deep enough to detect, on average, emission out to the optical radius $R_{25}$, in addition to providing generous sky coverage for excellent measurements of the background/foreground sky.  The removal of asteroids and detector artifacts is enabled via a second mapping that typically occurs a few days after the first visit.  The Basic Calibrated Data images have been processed with the standard SINGS IRAC pipeline.  The final pixel scale is $\sim$0\farcs75.  

All of the 24, 70, and 160\m\ images are being obtained in the MIPS scan mode (Rieke et al. 2004).  As with the IRAC imaging strategy, the MIPS maps are large enough to detect emission out to the optical radius, there is sufficient sky coverage, and asteroid removal is enabled via two separate passes at each source.  The final pixel scales are 0\farcs75, 3\farcs0, and 6\farcs0 at 24, 70, and 160\m, respectively.  These pixel scales allow us to finely sample the point spread function at each wavelength and were chosen to match (or be an integer multiple of) the IRAC pixel scale.
The MIPS data were processed using the MIPS Instrument Team Data Analysis Tool (Gordon et al. 2005).  Note that the MIPS data processing has evolved since our validation work on NGC~7331; removal of stimflash latent images in the 70\m\ data were only performed for NGC~7331, for example.  Further details of the MIPS data reduction procedures 
can be found in Bendo et al. (2005).


Foreground stars (and obvious background galaxies) were edited from each IRAC and MIPS image before flux extraction was performed.  Foreground stars were identified for this process by examining \Spitzercolord\ and \Spitzercolorc\ color images.  Global flux densities are extracted from apertures that cover the entire optical disk (see Table~\ref{tab:fluxes}).  The same apertures are used to measure flux densities in all MIPS and IRAC wavebands.  Sky subtraction was carried out through the use of multiple sky apertures placed near the source, yet do not overlap with the faintest isophotes visible from the galaxy.  Statistical uncertainties related to sky subtraction are usually less than 1\%, but can be appreciable (tens of percent) for faint sources.  Systematics in the IRAC calibration result in fluxes uncertain at the 10\% level (see 
Table~\ref{tab:calibration}).  
The uncertainties in the MIPS fluxes are also dominated by the uncertainties in the calibration: 10\% at 24\m\ and 20\% at 70 and 160\m.  The uncertainties listed in Table~\ref{tab:fluxes} include both systematic and statistical uncertainties.  No aperture corrections have been applied to the IRAC fluxes listed in Table~\ref{tab:fluxes}, nor in the plots throughout this work.  Aperture corrections for extended sources are still unknown and none of our conclusions would rely on such small changes (estimated to be between $\sim$4\% and 25\% in Pahre et al. 2004).

Submillimeter maps have been obtained with SCUBA for 27 SINGS galaxies, including 25 of the \N_MIPSIRAC\ in the present sample (Holland et al. 1999).  Thirteen of the maps were taken 
by us in jiggle map mode, whereas 14 of the maps are archival jiggle or scan maps.  Depending on the galaxy size, the maps run from $\sim$3\arcmin\ to several arcminutes in extent.  The maps are large enough to capture essentially all of the submillimeter emission for the galaxies in this work except for those listed in Table~\ref{tab:submm} with aperture correction factors.
Based on the fractional MIPS flux contained within apertures similar to the submillimeter map sizes and assuming no far-infrared-submillimeter color gradients, for these galaxies it is likely that we have only detected 
45-90\% of the total submillimeter emission.
The submillimeter fluxes listed in Table~\ref{tab:submm} 
have been corrected by the multiplicative factors in the same table.
The data were processed with the SCUBA User Reduction Facility (Jenness \& Lightfoot 1998).  The data were first flatfielded and corrected for atmospheric extinction.  Contributions from noisy bolometers were then removed, followed by spike removal.  For the jiggle maps the background signal was subtracted using the signal from several bolometers along the edges of the maps (the SCUBA maps are large enough to include sky for all galaxies except for NGC~5033; the sky estimation for this galaxy is suspect).  
For the scan maps the signal baseline and sky background were removed and then the data were calibrated, grouped according to chop throw and angle, and regridded.  The data were then smoothed with a 9\arcsec\ Guassian beam and combined together using the remdbm procedure in SURF.  A residual background was subtracted by smoothing the map with a boxcar function 150-300\arcsec\ in size and then subtracting the smoothed map from the original.  Note that this step may additionally subtract some diffuse, extended emission.
Further details of the submillimeter data processing can be found in Bendo et al. (2005).  The data were calibrated using observations of submillimeter standards and regridded onto the sky plane.  The calibration of the 450\m\ and 850\m\ fluxes are uncertain at the 25\% and 15\% levels, respectively.



Archival near-infrared, mid-infrared, and far-infrared data are used to supplement the \Spitzer\ and \SCUBA\ data, where available.  The 2MASS near-infrared fluxes used in this study are the total fluxes listed in the NASA/IPAC Extragalactic Database and are obtained by extrapolating the near-infrared surface brightness profiles out to about five disk scale lengths (see Jarrett et al. 2003).  Fluxes from the 2MASS All-Sky Extended Source Catalog exist for all but the six faintest of the 75 SINGS galaxies.
For these six sources we have extracted 2MASS fluxes using the same apertures used to determine IRAC and MIPS fluxes.  Flux uncertainty is at the 5\% level.  Archival 6.8 and 15\m\ fluxes (ISOCAM) are taken from Dale et al. (2000) and Roussel et al. (2001) for 13 SINGS galaxies.  In addition, we have extracted 6.8 and 15\m\ fluxes from archival ISOCAM data for another eight sources.  
ISOCAM fluxes are typically $\sim$20\% uncertain, though this percentage can substantially rise for very faint sources like Holmberg~II.  Finally, \IRAS\ fluxes, accurate to $\sim$20\% and obtained from SCANPI and HIRES data, are available in at least one wavelength 
for all but three SINGS targets.

\hal\ imaging was obtained for NGC~3031, NGC~5194, and NGC~7331 at the Kitt Peak National Observatory 2.1~m telescope in March 2001 (NGC~3031 and NGC~5194) and November 2001 (NGC~7331).  Integrations were 1800~s for the narrowband imaging and 420~s for the R band imaging that was used for continuum subtraction.  Single pointings of the 10\farcm4 field of view (0\farcs3~pixel$^{-1}$) were sufficient to capture the \hal\ emission for NGC~7331, but mosaics were necessary to encompass the large angular diameters of NGC~3031 and NGC~5194.  Flux calibration was obtained through observations of spectrophotometric sources from Massey et al. (1998).  The data were processed using standard IRAF routines.  A correction for \NII\ in these relatively metal-rich galaxies was made assuming a uniform value of \NII/\hal$=0.6$ (Hoopes \& Walterbos 2003).  Fluxes were further corrected for internal extinction (assuming $A_V\sim$1~mag; e.g. Kennicutt 1998) and foreground Milky Way extinction (Schlegel, Finkbeiner, \& Davis 1998).  The overall uncertainty in the \hal\ fluxes, including systematics, is estimated to be 25\%.

\section {Results}
\label{sec:results}

\subsection {Global Infrared-Submillimeter Spectral Energy Distributions}
\label{sec:global}


Figures~\ref{fig:seds1}-\ref{fig:seds8} show the infrared data for each galaxy; \2MASS, \IRAS, \ISO, and \SCUBA\ data are shown where available.  Included in each plot are two model curves and their sum.  
The dashed curve is a dust-only model from the semi-empirical work of Dale \& Helou (2002) on normal star-forming galaxies.  The dust-only templates have the advantage of being parametrized by a single parameter family, essentially defined by the far-infrared color.  The $\alpha_{\rm SED}$ listed within each panel parametrizes the distribution of dust mass as a function of local heating intensity $U$
\begin{eqnarray}
\label{eq:dMdU}
             dM_{\rm dust}(U) &\propto& U^{-\alpha_{\rm SED}} \; dU,  \; \; \; \; \; \; 0.3 \leq U \leq 10^5.
\end{eqnarray}
The dust curve ($\alpha_{\rm SED}$) is computed for each galaxy via 10,000 Monte Carlo simulations of a least-squares fit to the MIPS data.  For each simulation we add a random component to the MIPS fluxes, Gaussian-scaled by the fluxes' uncertainties.  The average $\alpha_{\rm SED}$ and its dispersion is given in each panel.  We have likewise adopted a conceptually simple approach for the stellar template.  The dotted curve is the 900~Myr continuous star formation, solar metallicity, Salpeter IMF ($\alpha_{\rm IMF}=2.35$) curve from Vazquez \& Leitherer (2005) fitted to the 2MASS fluxes.  Though our current approach is mostly illustrative and avoids a more sophisticated suite of multi-parameter dust and stellar models, it is remarkable how well the data are generally fit by such a simple combination.  Note that for a few galaxies the far-infrared emission appears to peak at a wavelength that is slightly longer than that in the most quiescent dust model (i.e. $\alpha_{\rm SED}=4.00$).  The dust-only models will be updated, and a more refined approach will be pursued for the stellar profiles, once all the SINGS data are in hand.
Finally, we note that the mid-infrared data for NGC~1377 are quite discrepant from the model predicted by the far-infrared fluxes.  Roussel et al. (2003) suggest that NGC~1377 is in the initial throes of an intense starburst, and that this heavily extincted system is being observed at such an early stage of the star formation that it is yet to show any optical signatures or synchrotron radiation.  The unusual mid-infrared properties of NGC~1377 are studied in detail by Roussel et al. (in preparation).

Information on our collection of infrared spectra can be more compactly displayed in a single SED plot and in an infrared color-color diagram.  The examples presented in Figure~\ref{fig:seds_sample} portray the variety of spectral shapes in the SINGS sample.  Arbitrarily normalizing the spectra at 8.0\m\ emphasizes the significant dispersion in the sample's near-infrared/mid-infrared and far-infrared/mid-infrared colors.  The top panel of Figure~\ref{fig:global_colors} shows the (dust-only) mid-infrared color as a function of the far-infrared color.  The stellar contribution is subtracted from the 8.0 and 24\m\ measures via extrapolation of the 3.6\m\ emission, assuming the 3.6\m\ emission is completely stellar:
\begin{equation}
\label{eq:fnu8*nu}
  f_\nu(8.0\mu {\rm m})_{\rm dust} \; / \; f_\nu(24\mu {\rm m})_{\rm dust} \; = \; (f_\nu(8.0\mu {\rm m}) - \eta_\nu^{8*} f_\nu(3.6\mu {\rm m})) \; / \; (f_\nu(24\mu {\rm m}) - \eta_\nu^{24*} f_\nu(3.6\mu {\rm m}))
\end{equation}
where $\eta_\nu^{8*} = 0.232$ and $\eta_\nu^{24*} = 0.032$ are chosen such that the dust-only flux densities vanish for a standard stellar spectrum (e.g. Helou et al. 2004; Engelbracht et al. 2005).  Likewise, the ratio of 8.0\m\ dust emission to {\it total} dust emission is plotted in the bottom panel of Figure~\ref{fig:global_colors} using the relation
\begin{equation}
\label{eq:fnu8*}
  (\nu f_\nu(8.0\mu {\rm m}) - \eta^{8*} \nu f_\nu(3.6\mu {\rm m})) \; / \; f_{\rm dust}(3-1100\mu {\rm m})
\end{equation}
where $\eta^{8*}=0.232 \times 3.6/8.0$
and the denominator is obtained from MIPS fluxes using Equation~4 in Dale \& Helou (2002).  The solid and dotted lines in Figure~\ref{fig:global_colors} respectively indicate the flavors of SED models presented in Dale \& Helou (2002) and Dale et al. (2001), derived from the average global trends for a sample of normal star-forming galaxies observed by \ISO\ and \IRAS.  The solid curve incorporates an improved approach to modeling the far-infrared/submillimeter emission (Dale \& Helou 2002).  

Some of the low-metallicity objects lie dramatically below the canonical star-forming galaxy curves, consistent with the Engelbracht et al. (2005) empirical finding that low metallicity galaxies show relatively low 8.0\m\ emission.  
These low-metallicity galaxies drive the scatter in the ratio between the inferred dust-only flux at 8\m\ and the 8\m\ flux as predicted from the dust model fitted to the MIPS fluxes.  The dispersion in this ratio is 2.90 and 1.82~dex including and not including, respectively, the low-metallicity systems.  A likely reason for the relatively low 8\m\ flux for low-metallicity galaxies is a different balance between PAH destruction and formation within these systems (PAHs dominate the emission at 8\m; see Figures~8 and 12 in Li \& Draine 2001).  For example, if PAHs are formed by grain-surface chemistry in the interstellar medium (so the grain formation rate would vary as the metallicity squared), and destroyed by gas-phase processes (sputtering in shock waves, or photodestruction) one might expect lower PAH/dust ratios in low metallicity systems.   Finally, we note that
the SINGS sample 
fills in more of the quiescent portion of the color-color space (e.g. small values of \Spitzercolora), and in fact some of the points lie beyond the most quiescent values predicted by the previous generation of models.
The SINGS sample spans a larger range of galaxy types, infrared/optical ratios, metallicities, and infrared colors than the sample from the \ISO\ Key Project on the Interstellar Medium of Normal Galaxies, so it is unsurprising that some of the data points in Figure~\ref{fig:global_colors} do not exactly match the models' predictions.

\subsection {Local Infrared-Submillimeter Spectral Energy Distributions}
\label{sec:local}

To properly execute a multi-wavelength investigation within galaxies, all the images were first convolved to match the point-spread function of the 160\m\ beam which has FWHM of 38\arcsec\ (see Engelbracht et al. 2004).  At the distances to NGC~3031, NGC~5194, and NGC~7331, 38\arcsec\ corresponds to spatial scales of $\sim$0.6, 1.5, and 2.8~kpc, respectively.  After the data were all smoothed to the same resolution, sky-subtracted flux densities were extracted from the 170$+$ various regions within each galaxy (Figures~\ref{fig:apertures1} and \ref{fig:apertures2}).

Figure~\ref{fig:local_colors} shows the mid-infrared color as a function of the far-infrared color for the various regions in these three galaxies.  
In general, the observed SEDs are indicative of more intense radiation fields for regions closer to the nucleus and arms, with larger \Spitzercolora\ ratios due to hotter large grains and smaller \Spitzercolorc\ ratios due to elevated small grain emission at 24\m.  The relatively low values for \Spitzercolorc\ near the nuclei and arm regions are unlikely explained by a diminution of the 8.0\m\ flux as a result of the destruction of polycyclic aromatic hydrocarbons, since the radiation fields suggested by \Spitzercolora\ within these galaxies are fairly mild.  This issue is addressed in Figure~\ref{fig:local_dust}.  The overall constancy of the 8.0\m-to-total infrared ratio in Figure~\ref{fig:local_dust} supports the conclusion that the relatively low values of \Spitzercolorc\ for high \Spitzercolora\ are due to elevated 24\m\ emission and not a reduction in the PAH emission.  Another interesting facet of the data plotted in Figure~\ref{fig:local_dust} is the comparatively small variation in 8.0\m-to-total dust emission: the typical scatter within each galaxy is $\sim$15\% even as \Spitzercolora\ varies by a factor of four.  This result reinforces the notion that the observed variations in \Spitzercolorc\ do not arise from variations in 8.0\m-to-total dust emission, but rather from variations in 24\m-to-total dust emission.

Another obvious feature to Figure~\ref{fig:local_colors} is the substantial variations seen within a given galaxy's disk.  The \Spitzercolorc\ and \Spitzercolora\ colors vary by a factor of two or more within each disk, and the dispersion is larger the closer the galaxy.  NGC~3031 shows the largest scatter because in this galaxy we are sampling the smallest spatial scales and thus the widest variety of environments.  Moreover, some of the local \Spitzercolora\ values are smaller than that predicted from the globally-integrated properties of the most quiescent galaxies studied by {\it ISO} and {\it IRAS} (i.e. the leftmost values of the model curves displayed in Figure~\ref{fig:local_colors}).  In other words, many of the local regions within these galaxies are more quiescent than what is seen for the global emission in normal star-forming galaxies.  This is not unexpected, as globally-integrated emission will inevitably encompass some hot spots, raising the overall average \Spitzercolora\ ratio above the typical cirrus value.

As alluded to above, the star formation rate is conceptually linked to the shape of the far-infrared spectrum---regions near enhanced star formation should have more intense interstellar radiation fields, and thus the large grains in such regions should be hotter than their counterparts in the diffuse interstellar medium.  However, before the advent of the {\it Spitzer Space Telescope} and its superior angular resolving power, it was difficult to explore correlations between the far-infrared color and the local star formation rate in all but the closest of galaxies (e.g. Xu \& Helou 1994; Xu \& Helou 1996).  Figure~\ref{fig:sfr} shows the far-infrared color as measured by \Spitzer\ versus the local star formation rate 
for NGC~3031, NGC~5194, and NGC~7331.  Star formation rates were computed using the relation presented in Kennicutt (1998): SFR ($M_\odot$~year$^{-1})=7.9 \times 10^{-42}~L$(\hal)~(ergs~s$^{-1}$).  The two quantities are clearly correlated; a non-parametric ranking analysis indicates a correlation at the $4-7\sigma$ level.  The ratio \Spitzercolore\ is likewise correlated with the local star formation rate, and \Spitzercolorb\ much less so.  These results suggest that both 24 and 70\m\ trace warmer dust phases, consistent with morphological studies carried out by Hinz et al. (2004) for M~33, Helou et al. (2004) for NGC~300, Gordon et al. (2004), and Rieke et al. for M~31 (private communication).  A large fraction of the 70\m\ emission may come from the same 
grains
that are thought to produce the 24\m\ emission (e.g. D\'{e}sert, Boulanger, \& Puget  1990), grains that are particularly sensitive to the effects of local star formation. 

\subsection{Implications for High-Redshift Observations: A Cautionary Tale}

One of the long-range goals of the SINGS project is to provide a reference set of empirical SED templates of nearby galaxies and regions within these galaxies that cover the full range of properties and interstellar environments found locally.  These templates can then be applied in turn to model the observed redshifted SEDs of galaxies observed at cosmological distances.  Such templates are already available from \ISO\ and \IRAS\ observations, such as the \ISO\ Key Project sample (e.g. Dale \& Helou 2002).  The SINGS sample is more diverse than the Key Project sample (metallicity, colors, morphology, etc.), and will eventually provide a definitive set of reference SEDs for such applications.  The preliminary data presented here already provide some useful guidance (and warnings) for interpreting infrared observations of high-redshift galaxies from 24\m\ or 8.0\m\ data alone (or together).

As a guide for \Spitzer-based cosmological studies, we show in Figure~\ref{fig:fir_z} how the \Spitzercolorf\ and the (dust-only) bolometric-to-24\m\ ratios would change as a function of redshift for a variety of SINGS-like objects.  The tracks are pinned at redshift zero according to each source's observed infrared properties, and traced out to higher redshifts using the model of Dale \& Helou (2002).  Figure~\ref{fig:fir_z} shows that SED variations remain a considerable source of systematic uncertainty even when the redshift is known.  Given the range of infrared spectral energy distributions observed for this portion of the SINGS galaxy sample, extrapolating the total infrared luminosity from the 24\m\ emission alone would yield uncertainties at the factor-of-five level.  This level of uncertainty can be reduced with some {\it a priori} constraints on the types and the redshifts of sources being studied (e.g. Papovich \& Bell 2002), as well as information on the infrared colors (e.g. \Spitzercolorc\ ratio or their redshifted equivalents).

However, for many current applications even the redshifted 24\m\ emission is inaccessible, and one must rely on the {\it observed} 24\m\ flux, which at $z = 2$ corresponds to 8.0\m\ in the rest frame.   An in-depth analysis of the application of this index at high redshift will be presented in a separate paper (Smith et al. 2005), but in the meantime we make some preliminary assessments based on the SED data shown in Figures~\ref{fig:global_colors} and \ref{fig:local_dust}.  Within individual large, metal-rich galaxies such as NGC~3031, NGC~5194, and NGC~7331 the ratio of 8.0\m-to-total infrared luminosity has a surprisingly low dispersion, less than a factor of two (maximum to minimum) within individual galaxies and hardly more among the three galaxies plotted in Figure~\ref{fig:local_dust}.  However if one considers a much larger galaxy sample (Figure~\ref{fig:global_colors}), the total range in 8.0\m-to-total infrared luminosity ratios swells to more than a factor of twenty.  Much of this large scatter can be attributed to a few metal-poor dwarf galaxies, in which the 8.0\m\ PAH emission is quenched.  If one excludes the metal-poor dwarfs the total range of 8.0\m-to-total ratios shrinks to a factor of 10, and this probably should be taken as a minimum systematic uncertainty in total infrared luminosities and star formation rates inferred from rest-frame 8.0\m\ fluxes alone (also see Calzetti et al. 2005).  If the metal abundance of the population cannot be inferred {\it a priori} then the uncertainties may be considerably higher.

\section {Summary}

New \Spitzer\ imaging data are coupled with data from \2MASS, \ISO, \IRAS, and \SCUBA\ to investigate the properties of the 1-850\m\ SEDs for \N_MIPSIRAC\ nearby galaxies from the SINGS sample.  The various shapes of the globally-integrated dust continuum generally follow the trends expected from the previous generation of infrared dust models that are based on \ISO\ and \IRAS\ observations of normal star-forming galaxies from the \ISO\ Key Project sample.  However, the SINGS sample is more diverse than the Key Project sample (metallicity, colors, morphology, etc), so it is perhaps not surprising that 
the SINGS sample shows larger variations in infrared colors.  

We also take advantage of \Spitzer's unprecedented angular resolving power to explore over 170 regions within NGC~3031, NGC~5194, and NGC~7331.  All three galaxies exhibit interesting trends in their infrared emission over the various environments sampled in this study.  Within each of these galaxies we find a wide variety of spectral shapes, especially for NGC~3031, the closest of the three targets and thus the galaxy for which we are able to sample the smallest spatial scales.  In particular we notice fairly constant ratios of dust-only 8.0\m\ emission with respect to both bolometric dust emission, over large dynamic ranges of dust temperature as measured by \Spitzercolora.  The relative contribution from PAHs does not appear to significantly vary on kiloparsec scales, suggesting that the variations observed in \Spitzercolorc\ are largely driven by variations in the local 24\m\ emission.  The emission from galaxies at 24\m, along with that at 70\m, traces 
grains
that are particularly sensitive to changes in the local interstellar radiation field and thus can be used as proxies for gauging the strength of the local star formation (e.g. Calzetti et al. 2005).

This dataset on a wide-ranging sample of local galaxies is useful for estimating how accurately one can predict the total infrared luminosity (and star formation rate) from just the 8.0 or 24\m\ data alone.  Though the 8.0\m-to-total infrared ratio varies by less than a factor of two within NGC~3031, NGC~5194, and NGC~7331, the global ratio varies by more than a factor of 20 among our diverse sample of galaxies.  The 24\m-to-total infrared ratio likewise spans a large range in the SINGS sample (factor of 5).  In conjunction with models of infrared SEDs for normal star-forming galaxies, we predict that the {\it observed} 24\m-to-total infrared ratio (e.g. rest-frame 8.0\m-to-total at $z=2$) maintains at least this dispersion at most epochs back to a redshift of four.

As mentioned above, the characteristics of the infrared SEDs as seen by \Spitzer\ are broadly consistent with normal galaxy models developed based on \ISO\ and \IRAS\ data, but with perhaps more significant dispersion.  Once the complete SINGS dataset is in hand, the suite of empirical SED models referenced here will be updated according to the diversity of spectral shapes exhibited by the SINGS project.  A more generally comprehensive and quantitative analysis will be pursued as well.  For example, how much more dispersion in infrared colors is seen within galaxies versus between galaxies?  How do the global spectral shapes depend on metallicity in this sample?  Are the infrared surface brightnesses correlated with the far-infrared color or galaxy type?  Data from all 75 galaxies in the diverse SINGS sample will help to answer these questions.

\acknowledgements 
We acknowledge E. Xilouris for helpful input.  Support for this work, part of the {\it Spitzer Space Telescope} Legacy Science Program, was provided by NASA through Contract Number 1224769 issued by the Jet Propulsion Laboratory, California Institute of Technology under NASA contract 1407.  This research has made use of the NASA/IPAC Extragalactic Database which is operated by JPL/Caltech, under contract with NASA.  This publication makes use of data products from the Two Micron All Sky Survey, which is a joint project of the University of Massachusetts and the Infrared Processing and Analysis Center/California Institute of Technology, funded by the National Aeronautics and Space Administration and the National Science Foundation.


\begin{deluxetable}{lrrrrrrr}
\def\a{\tablenotemark{a}}
\def\b{\tablenotemark{b}}

\def\p{$\pm$}
\tabletypesize{\scriptsize}
\tablenum{1}
\label{tab:fluxes}
\tablecaption{Infrared Flux Densities~\a\b}
\tablewidth{0pc}
\tablehead{
\colhead{Galaxy} &
\colhead{3.6\m} &
\colhead{4.5\m} &
\colhead{5.8\m} &
\colhead{8.0\m} &
\colhead{24\m} &
\colhead{70\m} &
\colhead{160\m} 
\\
\colhead{} & 
\colhead{(Jy)} &
\colhead{(Jy)} &
\colhead{(Jy)} &
\colhead{(Jy)} &
\colhead{(Jy)} &
\colhead{(Jy)} &
\colhead{(Jy)} 
}
\startdata
NGC~0024    &\nodata          &\nodata           &\nodata           &\nodata           &   0.13 \p   0.01 &   1.89 \p   0.39 &   6.723\p   1.36 \cr
NGC~0337    &   0.11\p   0.01 &   0.070\p   0.007&   0.17 \p   0.02 &   0.50 \p   0.05 &   0.65 \p   0.07 &   8.83 \p   1.77 &  18.30 \p   3.69 \cr
NGC~0584    &   0.40\p   0.04 &   0.23 \p   0.02 &   0.22 \p   0.02 &   0.15 \p   0.02 &   0.05 \p   0.006&   0.15 \p   0.07 &   1.02 \p   0.40 \cr
NGC~0628    &   0.95\p   0.10 &   0.58 \p   0.06 &   1.50 \p   0.15 &   3.62 \p   0.36 &   3.08 \p   0.31 &  29.73 \p   5.95 & 116.64 \p  23.34 \cr
NGC~0855    &   0.046\p  0.005&   0.029\p   0.003&   0.022\p   0.003&   0.060\p   0.006&   0.082\p   0.008&   1.38 \p   0.28 &   2.09 \p   0.44 \cr
NGC~0925    &\nodata          &\nodata           &\nodata           &\nodata           &   0.90 \p   0.09 &  12.20 \p   2.45 &  39.52 \p   7.95 \cr
NGC~1097    &   1.36\p   0.14 &   0.85 \p   0.09 &   1.89 \p   0.19 &   4.29 \p   0.43 &   6.40 \p   0.64 &  43.40 \p   8.68 & 144.63 \p  28.93 \cr
NGC~1266    &   0.059\p  0.006&   0.044\p   0.004&   0.066\p   0.007&   0.12 \p   0.01 &   0.84 \p   0.08 &   9.64 \p   1.93 &   9.14 \p   1.84 \cr
NGC~1291    &   2.32\p   0.23 &   1.35 \p   0.14 &   1.25 \p   0.13 &   0.86 \p   0.09 &   0.44 \p   0.05 &   5.41 \p   1.09 &  28.47 \p   5.75 \cr
NGC~1316    &   2.72\p   0.27 &   1.63 \p   0.16 &   1.46 \p   0.15 &   0.75 \p   0.08 &   0.36 \p   0.04 &   4.22 \p   0.85 &   9.66 \p   1.94 \cr
NGC~1377    &   0.062\p  0.006&   0.090\p   0.009&   0.32 \p   0.03 &   0.55 \p   0.06 &   1.73 \p   0.17 &   4.76 \p   0.95 &   2.91 \p   0.60 \cr
NGC~1404    &   0.80\p    0.08&   0.46 \p   0.05 &   0.42 \p   0.04 &   0.21 \p   0.02 &   0.083\p   0.009&   0.15 \p   0.09 &   0.31 \p   0.18 \cr
NGC~1512    &   0.43\p    0.04&   0.26 \p   0.03 &   0.34 \p   0.03 &   0.59 \p   0.06 &   0.42 \p   0.04 &   5.40 \p   1.08 &  21.85 \p   4.38 \cr
NGC~1566    &   0.82\p    0.08&   0.51 \p   0.05 &   1.16 \p   0.12 &   2.84 \p   0.28 &   2.65 \p   0.27 &  27.82 \p   5.57 &  95.26 \p  19.05 \cr
NGC~1705    &   0.028\p  0.003&   0.019\p   0.002&   0.012\p   0.002&   0.022\p   0.002&   0.052\p   0.005&   1.09 \p   0.22 &   1.20 \p   0.25 \cr
NGC~2403    &   2.06\p    0.21&   1.38 \p   0.14 &   2.79 \p   0.28 &   5.53 \p   0.55 &   5.64 \p   0.56 &  75.58 \p  15.12 & 231.56 \p  46.32 \cr
Holmberg~II &   0.078\p  0.008&   0.060\p   0.006&   0.039\p   0.005&   0.032\p   0.005&   0.17 \p   0.02 &   3.18 \p   0.64 &   4.05 \p   0.87 \cr
M81~Dwarf~A &   0.002\p  0.001&   0.001\p   0.001&  $<$0.001        &  $<$0.001        &\nodata           &\nodata           &\nodata           \cr
DDO~053     &   0.006\p  0.001&   0.005\p   0.001&   0.003\p   0.001&   0.010\p   0.001&   0.028\p   0.003&   0.31 \p   0.07 &   0.32 \p   0.11 \cr
NGC~2798    &   0.13\p   0.01 &   0.086\p   0.009&   0.32 \p   0.03 &   0.84 \p   0.08 &   2.51 \p   0.25 &  14.70 \p   2.94 &  18.45 \p   3.69 \cr
NGC~2841    &   1.39\p   0.14 &   0.80 \p   0.08 &   0.85 \p   0.09 &   1.56 \p   0.16 &   0.88 \p   0.09 &   8.66 \p   1.74 &  54.87 \p  10.98 \cr
NGC~2915    &\nodata          &\nodata           &\nodata           &\nodata           &   0.058\p   0.006&   1.09 \p   0.22 &   1.09 \p   0.30 \cr
Holmberg~I  &   0.013\p  0.001&   0.008\p   0.001&   0.009\p   0.002&   0.010\p   0.002&   0.013\p   0.004&   0.33 \p   0.12 &   0.76 \p   0.23 \cr
NGC~2976    &   0.47\p   0.05 &   0.30 \p   0.03 &   0.64 \p   0.06 &   1.36 \p   0.14 &   1.33 \p   0.13 &  16.99 \p   3.40 &  46.81 \p   9.40 \cr
NGC~3049    &   0.044\p  0.004&   0.029\p   0.003&   0.078\p   0.008&   0.18 \p   0.02 &   0.41 \p   0.04 &   2.27 \p   0.46 &   4.05 \p   0.82 \cr
NGC~3031    &  11.87\p   1.19 &   6.90 \p   0.69 &   7.90 \p   0.79 &  10.78 \p   1.08 &   4.94 \p   0.49 &  74.37 \p  14.88 & 347.10 \p  69.43 \cr
Holmberg~IX &   0.008\p  0.001&   0.004\p   0.001&  $<$0.006        &  $<$0.006        &\nodata           &\nodata           &\nodata           \cr
M81~Dwarf~B &   0.005\p  0.001&   0.004\p   0.001&   0.003\p   0.001&   0.003\p   0.001&   0.008\p   0.001&   0.12 \p   0.03 &   0.21 \p   0.14 \cr
NGC~3190    &   0.41\p   0.04 &   0.25 \p   0.03 &   0.30 \p   0.03 &   0.44 \p   0.04 &   0.26 \p   0.03 &   4.34 \p   0.87 &  13.19 \p   2.65 \cr
NGC~3198    &   0.30\p   0.03 &   0.18 \p   0.02 &   0.42 \p   0.04 &   0.92 \p   0.09 &   1.03 \p   0.10 &   8.68 \p   1.74 &  34.96 \p   7.00 \cr
IC~2574     &   0.17\p   0.02 &   0.096\p   0.01 &   0.084\p   0.008&   0.089\p   0.009&   0.27 \p   0.03 &   4.61 \p   0.92 &  10.31 \p   2.12 \cr
NGC~3265    &\nodata          &\nodata           &\nodata           &\nodata           &   0.28 \p   0.03 &   2.05 \p   0.42 &   2.35 \p   0.49 \cr
Markarian~33&   0.029\p  0.003&   0.020\p   0.002&   0.063\p   0.006&   0.17 \p   0.02 &   0.82 \p   0.08 &   3.34 \p   0.67 &   3.46 \p   0.71 \cr
NGC~3351    &   0.89\p   0.09 &   0.55 \p   0.06 &   0.93 \p   0.09 &   1.80 \p   0.18 &   2.40 \p   0.24 &  16.42 \p   3.29 &  59.72 \p  11.95 \cr
NGC~3521    &   2.23\p   0.22 &   1.44 \p   0.14 &   3.29 \p   0.33 &   8.36 \p   0.84 &   5.36 \p   0.54 &  49.85 \p   9.97 & 206.65 \p  41.35 \cr
\enddata
\tablenotetext{a}{\footnotesize Flux uncertainties include both calibration and statistical uncertainties.  Calibration errors are 10\% at 3.6, 4.5, 5.8, 8.0, and 24\m, and 20\% at 70 and 160\m.}
\tablenotetext{b}{\footnotesize The `raw' IRAC flux densities are listed; i.e. they are not corrected for aperture effects.}
\end{deluxetable}


\begin{deluxetable}{lrrrrrrr}
\def\a{\tablenotemark{a}}
\def\b{\tablenotemark{b}}

\def\p{$\pm$}
\tabletypesize{\scriptsize}
\tablecaption{Infrared Flux Densities~\a\b (continued)}
\tablewidth{0pc}
\tablehead{
\colhead{Galaxy} &
\colhead{3.6\m} &
\colhead{4.5\m} &
\colhead{5.8\m} &
\colhead{8.0\m} &
\colhead{24\m} &
\colhead{70\m} &
\colhead{160\m} 
\\
\colhead{} & 
\colhead{(Jy)} &
\colhead{(Jy)} &
\colhead{(Jy)} &
\colhead{(Jy)} &
\colhead{(Jy)} &
\colhead{(Jy)} &
\colhead{(Jy)} 
}
\startdata
NGC~3621    &\nodata          &\nodata           &\nodata           &\nodata           &   3.30 \p   0.33 &  40.21 \p   8.04 & 126.15 \p  25.24 \cr
NGC~3627    &   2.05\p   0.21 &   1.32 \p   0.13 &   3.06 \p   0.31 &   7.50 \p   0.75 &   7.25 \p   0.73 &  68.92 \p  13.79 & 208.13 \p  41.63 \cr
NGC~3773    &   0.024\p  0.002&   0.015\p   0.002&   0.029\p   0.003&   0.061\p   0.006&   0.13 \p   0.01 &   1.22 \p   0.25 &   2.12 \p   0.48 \cr
NGC~3938    &   0.35\p   0.04 &   0.23 \p   0.02 &   0.52 \p   0.05 &   1.32 \p   0.13 &   1.05 \p   0.11 &  12.14 \p   2.43 &  46.78 \p   9.36 \cr
NGC~4125    &   0.70\p   0.07 &   0.39 \p   0.04 &   0.30 \p   0.03 &   0.19 \p   0.02 &   0.069\p   0.007&   0.86 \p   0.18 &   1.33 \p   0.30 \cr
NGC~4236    &\nodata          &\nodata           &\nodata           &\nodata           &   0.53 \p   0.05 &   7.08 \p   1.42 &  18.87 \p   3.85 \cr
NGC~4254    &   0.77\p   0.08 &   0.50 \p   0.05 &   1.89 \p   0.19 &   5.28 \p   0.53 &   4.09 \p   0.41 &  39.02 \p   7.80 & 131.79 \p  26.36 \cr
NGC~4321    &   1.04\p   0.10 &   0.68 \p   0.07 &   1.55 \p   0.16 &   3.88 \p   0.39 &   3.33 \p   0.33 &  32.28 \p   6.46 & 128.41 \p  25.68 \cr
NGC~4450    &   0.58\p   0.06 &   0.35 \p   0.03 &   0.32 \p   0.03 &   0.36 \p   0.04 &   0.19 \p   0.02 &   2.46 \p   0.50 &  13.73 \p   2.76 \cr
NGC~4536    &\nodata          &\nodata           &\nodata           &\nodata           &   3.37 \p   0.34 &  22.49 \p   4.50 &  54.39 \p  10.89 \cr
NGC~4552    &   0.91\p   0.09 &   0.51 \p   0.05 &   0.37 \p   0.04 &   0.23 \p   0.02 &   0.062\p   0.006&   0.097\p   0.04 &   0.41 \p   0.41 \cr
NGC~4559    &   0.39\p   0.04 &   0.25 \p   0.03 &   0.53 \p   0.05 &   1.13 \p   0.11 &   1.08 \p   0.11 &  14.32 \p   2.87 &  46.81 \p   9.37 \cr
NGC~4569    &   0.83\p   0.08 &   0.50 \p   0.05 &   0.75 \p   0.08 &   1.36 \p   0.14 &   1.41 \p   0.14 &   9.65 \p   1.93 &  38.21 \p   7.66 \cr
NGC~4579    &   0.95\p   0.10 &   0.55 \p   0.06 &   0.66 \p   0.07 &   0.97 \p   0.10 &   0.74 \p   0.07 &   8.21 \p   1.65 &  39.07 \p   7.82 \cr
NGC~4594    &   4.29\p   0.43 &   2.44 \p   0.24 &   2.18 \p   0.22 &   1.74 \p   0.17 &   0.65 \p   0.07 &   6.71 \p   1.36 &  36.84 \p   7.39 \cr
NGC~4631    &   1.38\p   0.14 &   0.89 \p   0.09 &   3.22 \p   0.32 &   7.89 \p   0.79 &   7.97 \p   0.80 &  98.78 \p  19.76 & 269.01 \p  53.80 \cr
NGC~4725    &   1.25\p   0.13 &   0.75 \p   0.08 &   0.97 \p   0.10 &   1.63 \p   0.16 &   0.81 \p   0.08 &   7.48 \p   1.50 &  53.42 \p  10.70 \cr
NGC~4736    &   3.95\p   0.40 &   2.46 \p   0.25 &   3.61 \p   0.36 &   6.97 \p   0.70 &   5.50 \p   0.55 &  69.89 \p  13.98 & 170.28 \p  34.06 \cr
DDO~154     &  0.0012\p 0.0004&  0.0008\p  0.0003& $<$0.0012        & $<$0.0009        &  0.0058\p   0.002&   0.043\p   0.03 &   0.26 \p   0.14 \cr
NGC~4826    &   2.76\p   0.28 &   1.67 \p   0.17 &   2.13 \p   0.21 &   3.16 \p   0.32 &   2.47 \p   0.25 &  35.68 \p   7.14 &  85.39 \p  17.09 \cr
DDO~165     &\nodata          &\nodata           &\nodata           &\nodata           &   0.011\p   0.003&   0.14 \p   0.05 &   0.27 \p   0.15 \cr
NGC~5033    &   0.70\p   0.07 &   0.50 \p   0.05 &   1.05 \p   0.11 &   2.59 \p   0.26 &   1.92 \p   0.19 &  21.50 \p   4.30 &  88.15 \p  17.63 \cr
NGC~5055    &   2.61\p   0.26 &   1.64 \p   0.16 &   3.47 \p   0.35 &   7.60 \p   0.76 &   5.59 \p   0.56 &  59.76 \p  11.95 & 286.34 \p  57.27 \cr
NGC~5194    &   2.91\p   0.29 &   1.90 \p   0.19 &   5.70 \p   0.57 &  14.33 \p   1.43 &  12.25 \p   1.23 & 131.36 \p  26.30 & 494.34 \p  98.99 \cr
NGC~5195    &   0.91\p   0.09 &   0.54 \p   0.05 &   0.56 \p   0.06 &   0.85 \p   0.09 &   1.31 \p   0.05 &  10.85 \p   2.17 &  12.34 \p   2.49 \cr
Tololo~89   &   0.041\p  0.004&   0.026\p   0.003&   0.017\p   0.002&   0.077\p   0.008&   0.25 \p   0.03 &   1.52 \p   0.31 &   2.69 \p   0.59 \cr
NGC~5408    &   0.056\p  0.006&   0.039\p   0.004&   0.050\p   0.005&   0.050\p   0.005&   0.42 \p   0.04 &   2.95 \p   0.59 &   2.21 \p   0.49 \cr
NGC~5474    &   0.11\p   0.01 &   0.085\p   0.009&   0.11 \p   0.01 &   0.15 \p   0.02 &   0.18 \p   0.02 &   3.17 \p   0.64 &   9.49 \p   1.92 \cr
NGC~5713    &   0.22\p   0.02 &   0.15 \p   0.02 &   0.35 \p   0.04 &   1.52 \p   0.15 &   2.28 \p   0.23 &  17.23 \p   3.45 &  34.77 \p   6.96 \cr
NGC~5866    &\nodata          &\nodata           &\nodata           &\nodata           &   0.20 \p   0.02 &   6.66 \p   1.33 &  16.53 \p   3.31 \cr
IC~4710     &   0.076\p  0.008&   0.049\p   0.005&   0.055\p   0.006&   0.086\p   0.009&   0.11 \p   0.01 &   1.97 \p   0.40 &   3.15 \p   0.67 \cr
NGC~6822    &   2.24\p   0.22 &   1.41 \p   0.14 &   1.91 \p   0.19 &   0.75 \p   0.08 &   2.51 \p   0.25 &  53.21 \p  10.65 & 136.22 \p  27.27 \cr
NGC~6946    &   3.47\p   0.35 &   2.26 \p   0.23 &   7.51 \p   0.75 &  18.49 \p   1.85 &  20.87 \p   2.09 & 177.89 \p  35.58 & 498.35 \p  99.71 \cr
NGC~7331    &   1.75 \p  0.17 &   1.08 \p   0.11 &   2.36 \p   0.24 &   5.40 \p   0.54 &   3.92 \p   0.39 &  56.49 \p  11.30 & 164.12 \p  32.89 \cr
NGC~7552    &   0.50 \p  0.05 &   0.38 \p   0.04 &   1.33 \p   0.13 &   3.63 \p   0.36 &  10.30 \p   1.03 &  45.40 \p   9.09 &  86.65 \p  17.34 \cr
NGC~7793    &   0.85 \p  0.08 &   0.50 \p   0.05 &   1.33 \p   0.13 &   2.49 \p   0.25 &   1.97 \p   0.20 &  29.86 \p   5.97 & 119.53 \p  23.91 \cr
\enddata
\tablenotetext{a}{\footnotesize Flux uncertainties include both calibration and statistical uncertainties.  Calibration errors are 10\% at 3.6, 4.5, 5.8, 8.0, and 24\m, and 20\% at 70 and 160\m.}
\tablenotetext{b}{\footnotesize The `raw' IRAC flux densities are listed; i.e. they are not corrected for aperture effects.}
\end{deluxetable}


\begin{deluxetable}{lc}
\def\a{\tablenotemark{a}}
\tabletypesize{\scriptsize}
\tablenum{2}
\label{tab:calibration}
\tablecaption{Flux Calibration Values}
\tablewidth{0pc}
\tablehead{
\colhead{Band} &
\colhead{Conversion\a} \\
\colhead{} & 
\colhead{(MJy~sr$^{-1}$ per DN~s$^{-1}$)} 
}
\startdata
IRAC 3.6\m&0.1125 \cr
IRAC 4.5\m&0.1375 \cr
IRAC 5.8\m&0.5913 \cr
IRAC 8.0\m&0.2008 \cr
MIPS 24\m &0.0439 \cr
MIPS 70\m &634.   \cr
MIPS 160\m&42.6   \cr
\enddata
\tablenotetext{a}{\footnotesize The values for NGC~7331 IRAC data are 0.111, 0.133, 0.583, 0.195.}
\end{deluxetable}



\begin{deluxetable}{lcccc}
\def\a{\tablenotemark{a}}
\tabletypesize{\scriptsize}
\def\p{$\pm$}
\tablenum{3}
\label{tab:submm}
\tablecaption{Submillimeter Fluxes and Aperture Correction Factors}
\tablewidth{0pc}
\tablehead{
\colhead{Galaxy} &
\colhead{450\m} &
\colhead{850\m} &
\colhead{450\m} &
\colhead{850\m} \\
\colhead{}     &
\colhead{(Jy)} &
\colhead{(Jy)} &
\colhead{Correction} &
\colhead{Correction} 
}
\startdata
NGC~0337    &\nodata        & 0.35\p0.05 & \nodata & \nodata \cr
NGC~1097    &\nodata        & 1.44\p0.78 & \nodata & 2.09    \cr
NGC~2798    &\nodata        & 0.19\p0.03 & \nodata & 1.08    \cr
NGC~2976    &\nodata        & 0.61\p0.24 & \nodata & 1.56    \cr
NGC~3190    &\nodata        & 0.19\p0.04 & \nodata & 1.12    \cr
Markarian~33&\nodata        & 0.04\p0.01 & \nodata & \nodata \cr
NGC~3521    &\nodata        & 2.11\p0.82 & \nodata & 1.56    \cr
NGC~3627    &\nodata        & 1.86\p0.70 & \nodata & 1.53    \cr
NGC~4254    &\nodata        & 1.01\p0.54 & \nodata & 2.06    \cr
NGC~4321    &\nodata        & 0.88\p0.49 & \nodata & 2.19    \cr
NGC~4536    &\nodata        & 0.42\p0.11 & \nodata & 1.30    \cr
NGC~4569    &\nodata        & 0.47\p0.08 & \nodata & 1.11    \cr
NGC~4579    &\nodata        & 0.44\p0.07 & \nodata & \nodata \cr
NGC~4594    &\nodata        & 0.37\p0.11 & \nodata & 1.33    \cr
NGC~4631    &  30.70\p10.02 & 5.73\p1.21 & 1.27    & 1.17    \cr
NGC~4736    &\nodata        & 1.54\p0.66 & \nodata & 1.67    \cr
NGC~4826    &\nodata        & 1.23\p0.31 & \nodata & 1.24    \cr
NGC~5033    &\nodata        & 1.10\p0.55 & \nodata & 1.93    \cr
NGC~5194    &\nodata        & 2.61\p0.39 & \nodata & \nodata \cr
NGC~5195    &\nodata        & 0.26\p0.04 & \nodata & \nodata \cr
NGC~5713    &\nodata        & 0.57\p0.12 & \nodata & 1.17    \cr
NGC~5866    &   0.79\p 0.20 & 0.14\p0.02 & \nodata & \nodata \cr
NGC~6946    &  18.53\p 4.63 & 2.98\p0.45 & \nodata & \nodata \cr
NGC~7331    &  20.56\p 8.10 & 2.11\p0.38 & 1.44    & 1.11    \cr
NGC~7552    &\nodata        & 0.80\p0.17 & \nodata & 1.17    \cr
\enddata
\end{deluxetable}

\clearpage

\begin{figure}
 \caption{The various local apertures selected for NGC~3031 (M~81) and NGC~5194 (M~51) overlaid on (unsmoothed) IRAC 8.0\m\ imaging.  The apertures are at least as large as the $\sim$38\arcsec\ FWHM of the 160\m\ point spread function, the point spread function to which all other images are smoothed for our analysis of local spectral energy distributions.  North is up, East is to the left.  Cyan$\rightarrow$nuclear; green$\rightarrow$inter-arm; blue$\rightarrow$arm; red$\rightarrow$inner disk; magenta$\rightarrow$outer disk.
}
 \label{fig:apertures1}
\end{figure}

\begin{figure}
 \caption{Similar to Figure~\ref{fig:apertures1} for NGC~7331.}
 \label{fig:apertures2}
\end{figure}

\begin{figure}
 \plotone{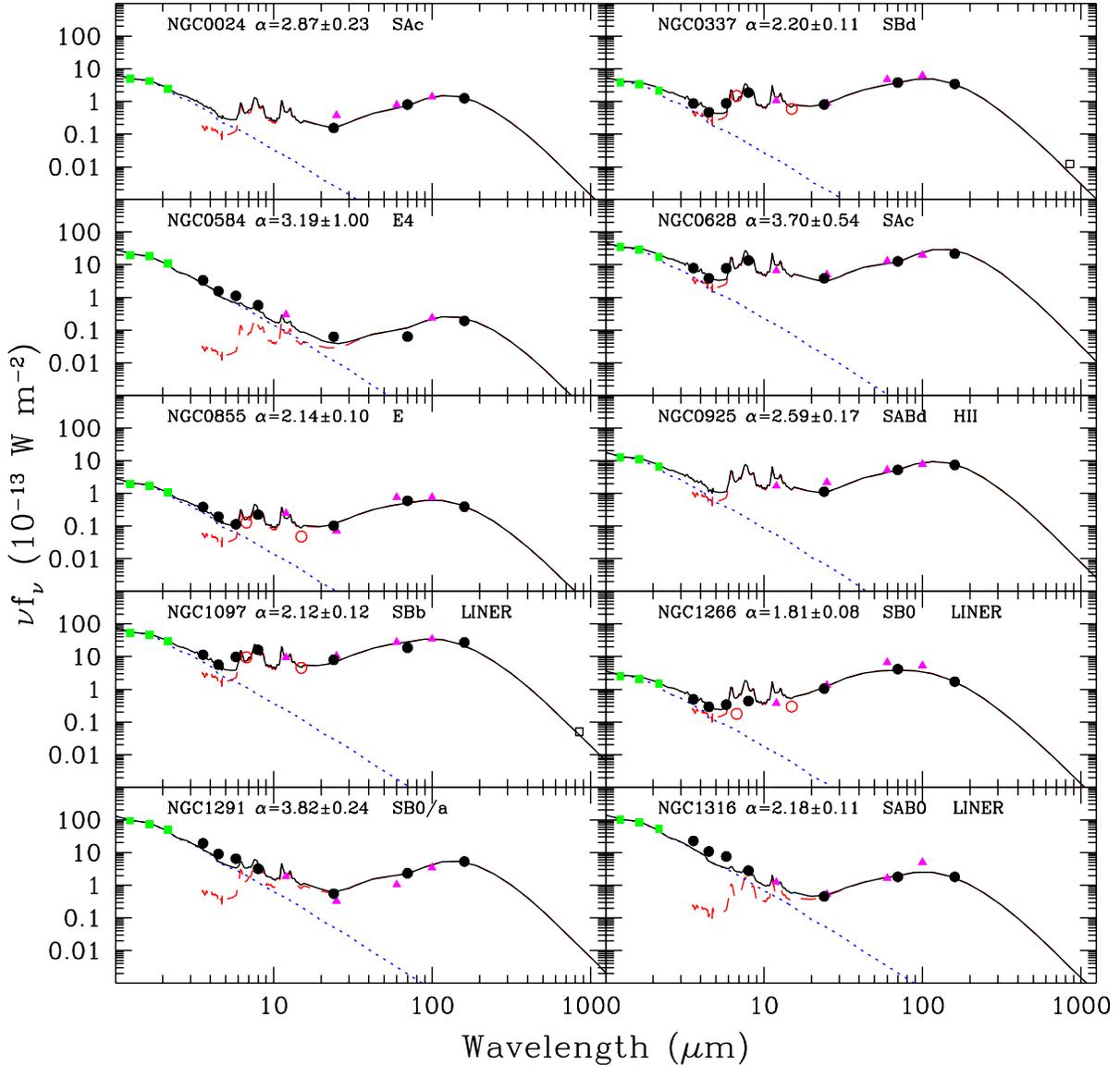}
 \caption{A presentation of the globally-integrated 1-850\m\ spectral energy distributions for 10 SINGS galaxies.  \2MASS, \Spitzer, \IRAS, \ISO, and \SCUBA\ data are represented by filled squares, filled circles, filled triangles, open circles, and open squares, respectively.  The solid curve is the sum of a dust (dashed) and a stellar (dotted) model.  The dust curve is a Dale \& Helou (2002) model fitted to the 24, 70, and 160\m\ fluxes; the $\alpha_{\rm SED}$ listed within each panel parametrizes the distribution of dust mass as a function of heating intensity, as described in Equation~\ref{eq:dMdU} and Dale \& Helou (2002).  The stellar curve is the 900~Myr continuous star formation, solar metallicity, Salpeter IMF ($\alpha_{\rm IMF}=2.35$) curve from Vazquez \& Leitherer (2005) fitted to the 2MASS data.}
 \label{fig:seds1}
\end{figure}

\begin{figure}
 \plotone{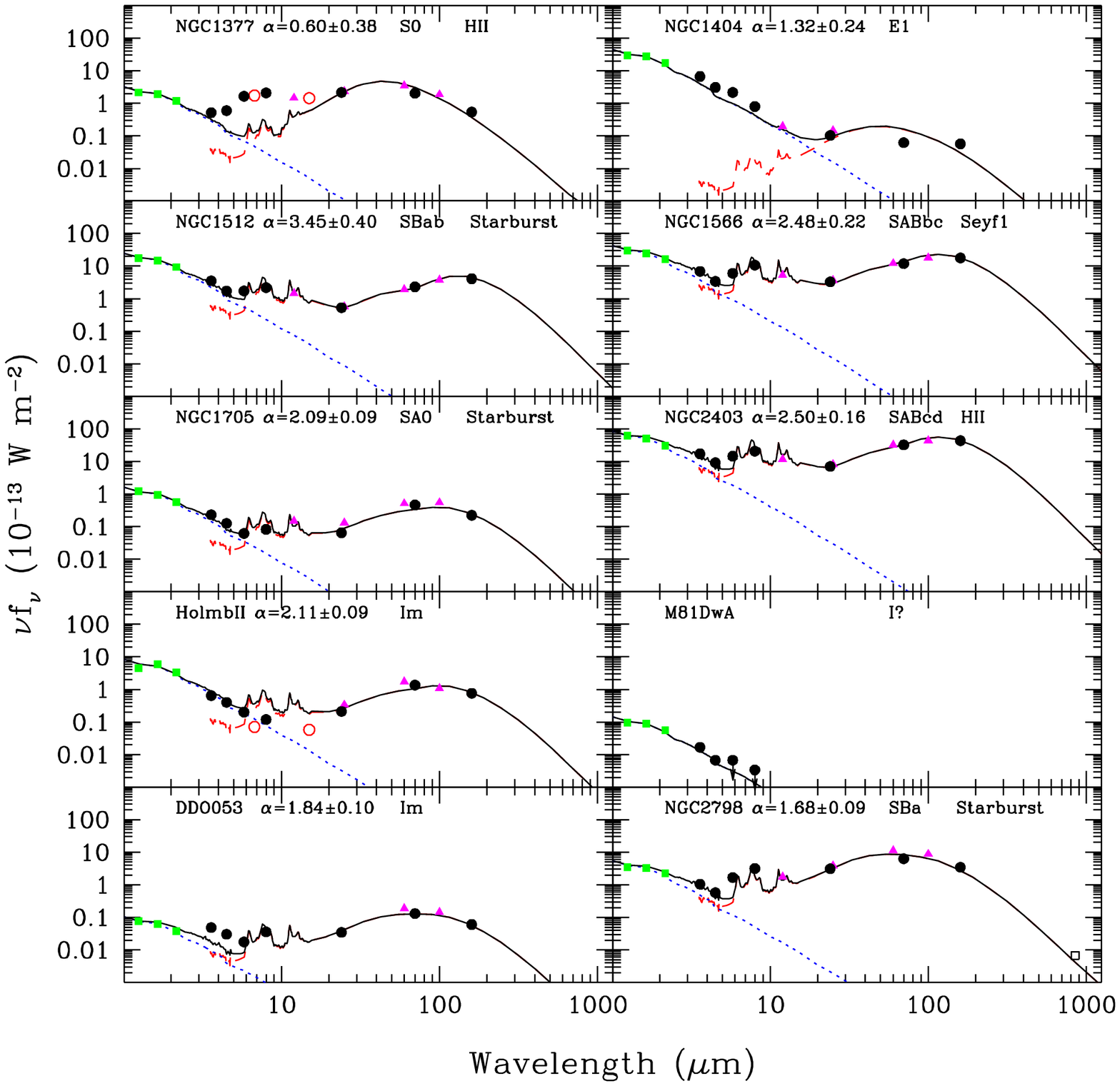}
 \caption{Similar to Figure~\ref{fig:seds1} for 10 more SINGS galaxies.}
 \label{fig:seds2}
\end{figure}

\begin{figure}
 \plotone{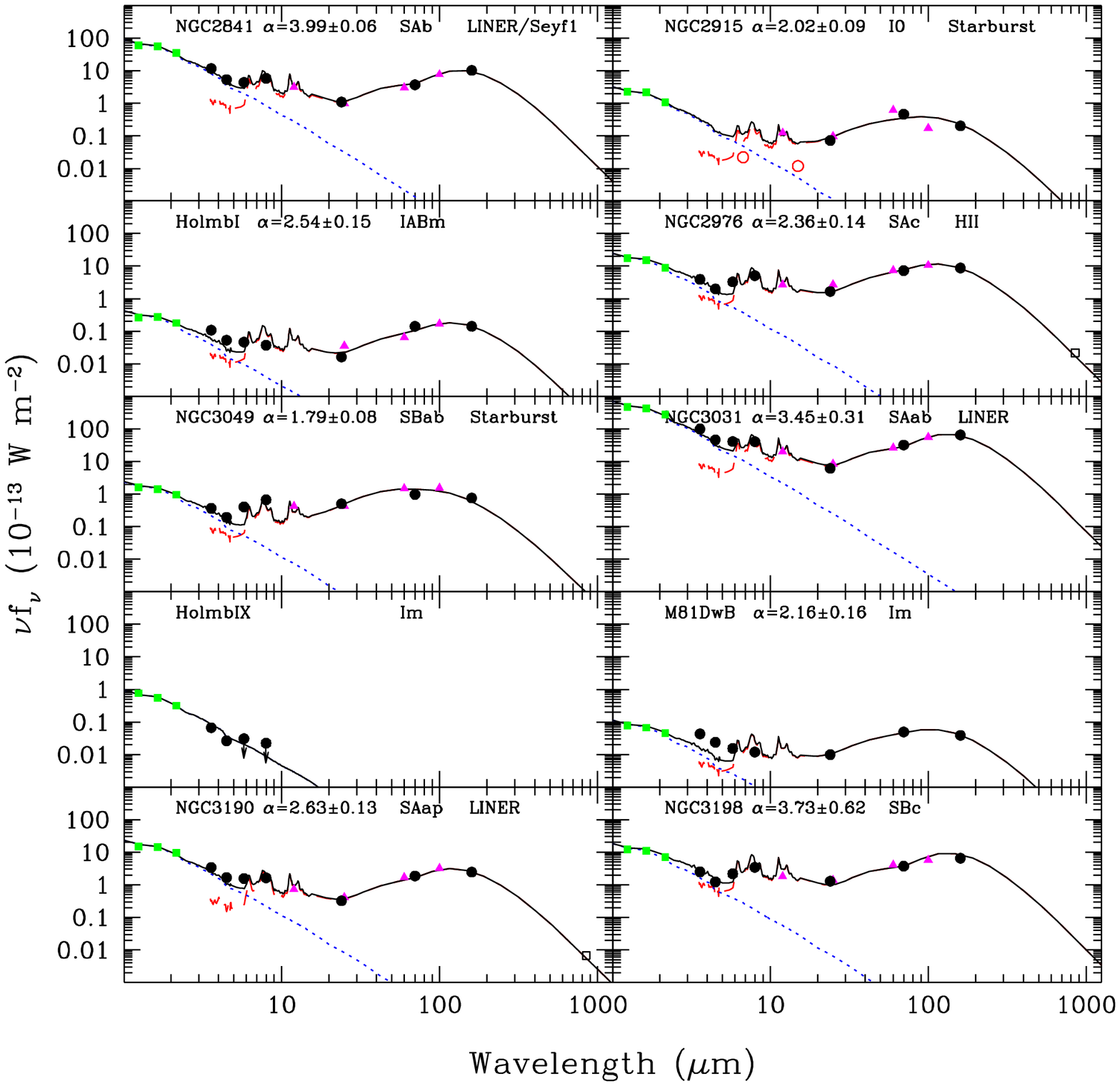}
 \caption{Similar to Figure~\ref{fig:seds1} for 10 more SINGS galaxies.}
 \label{fig:seds3}
\end{figure}

\begin{figure}
 \plotone{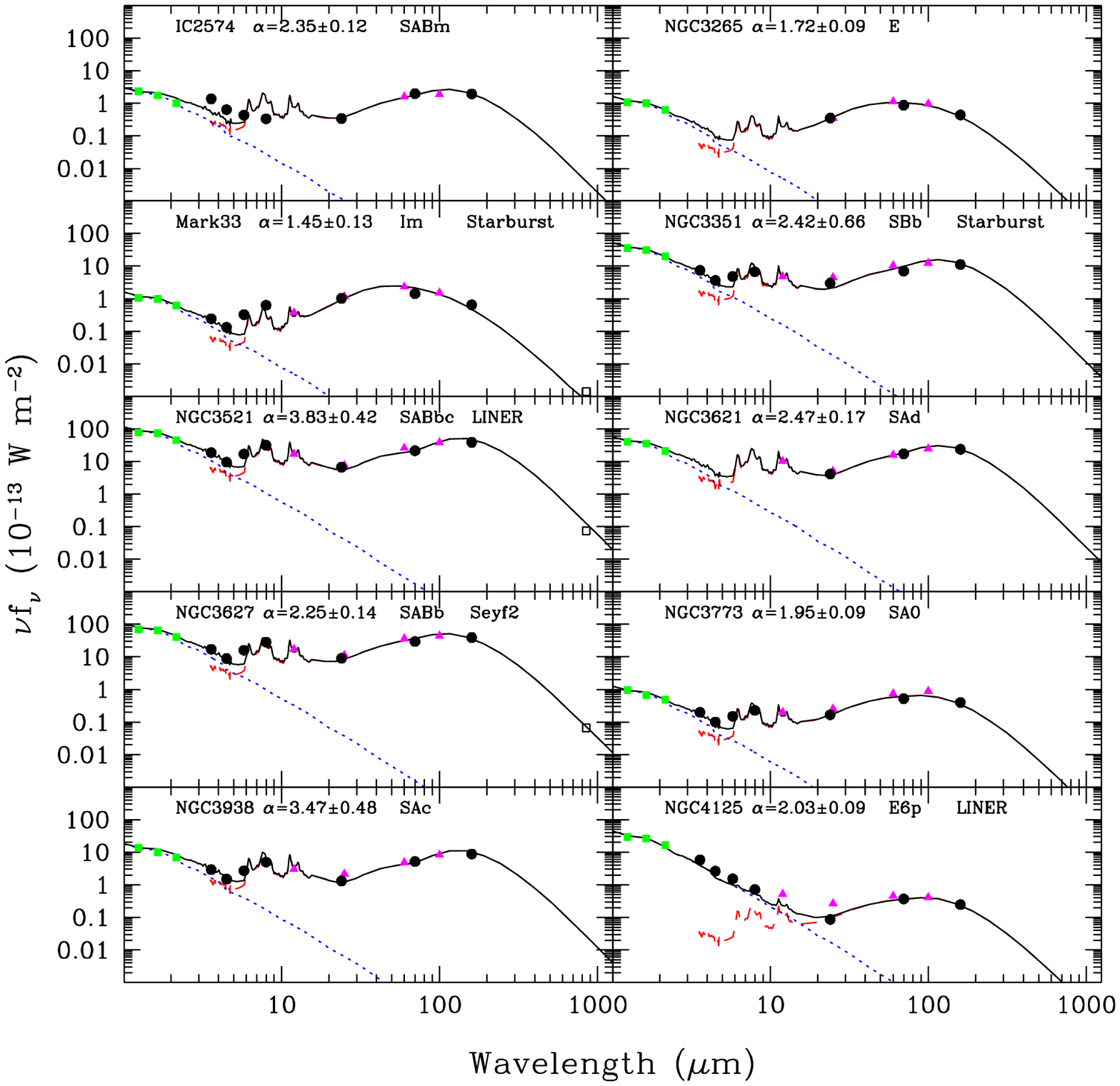}
 \caption{Similar to Figure~\ref{fig:seds1} for 10 more SINGS galaxies.}
 \label{fig:seds4}
\end{figure}

\begin{figure}
 \plotone{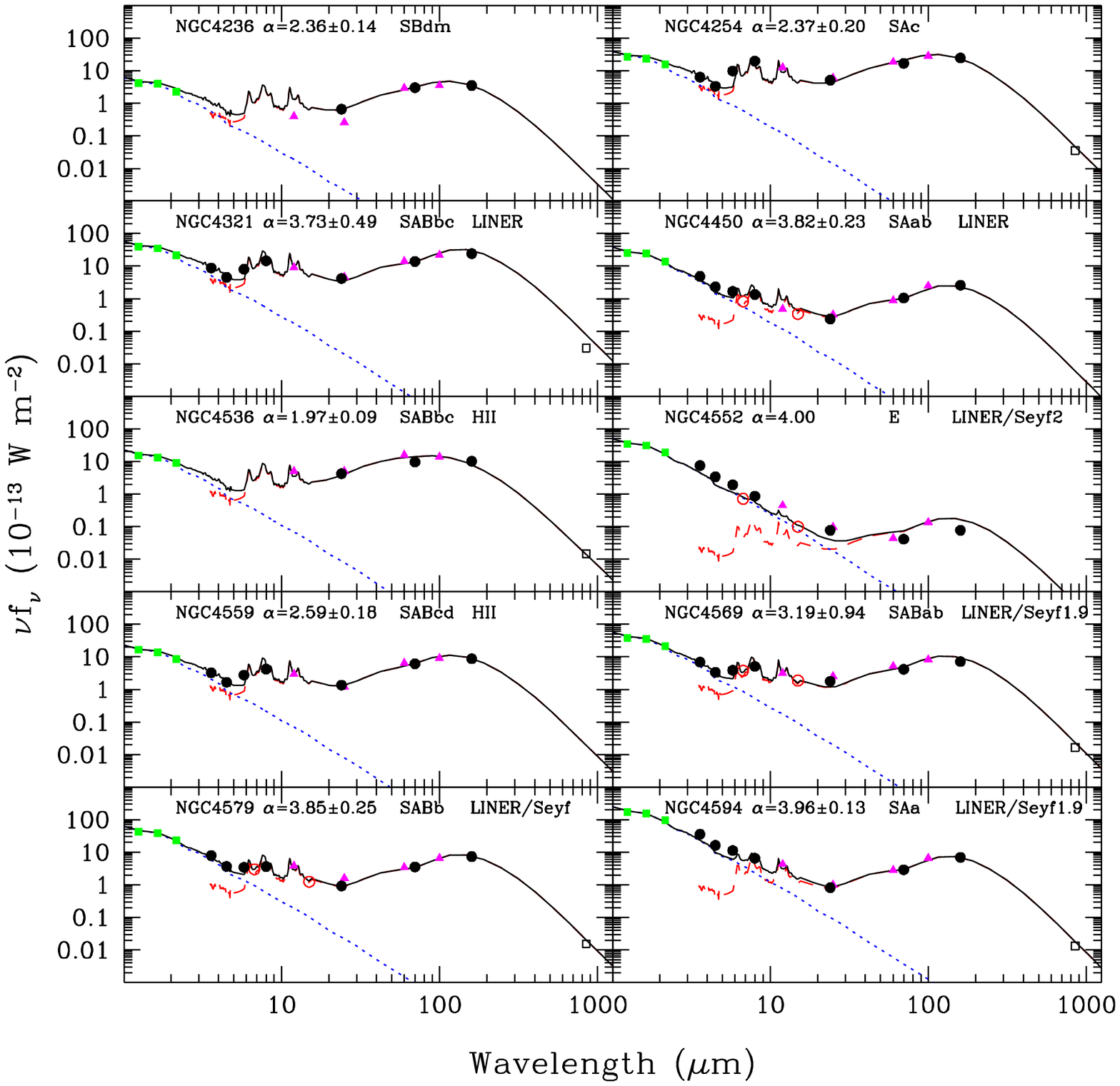}
 \caption{Similar to Figure~\ref{fig:seds1} for 10 more SINGS galaxies.}
 \label{fig:seds5}
\end{figure}

\begin{figure}
 \plotone{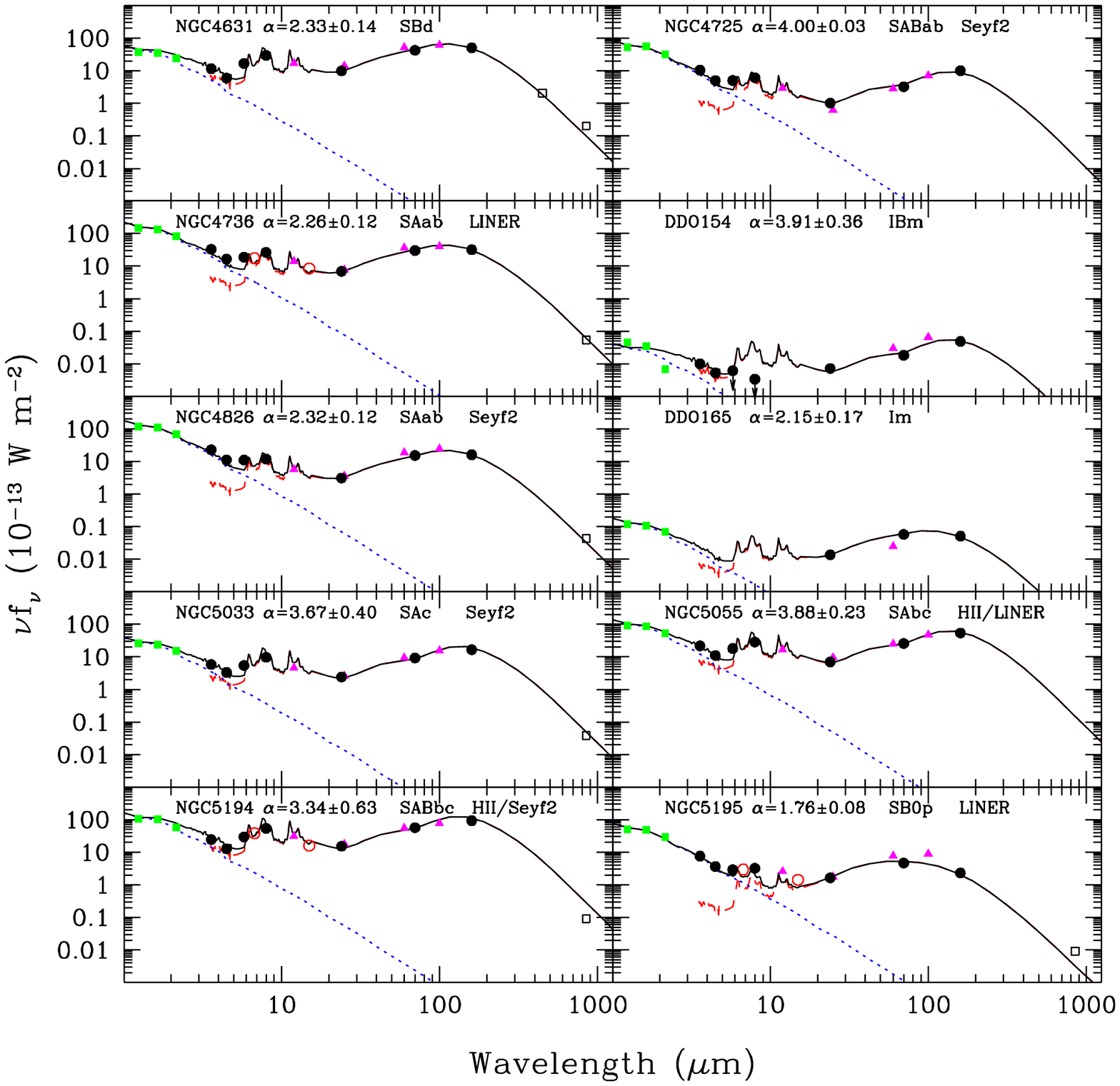}
 \caption{Similar to Figure~\ref{fig:seds1} for 10 more SINGS galaxies.}
 \label{fig:seds6}
\end{figure}

\begin{figure}
 \plotone{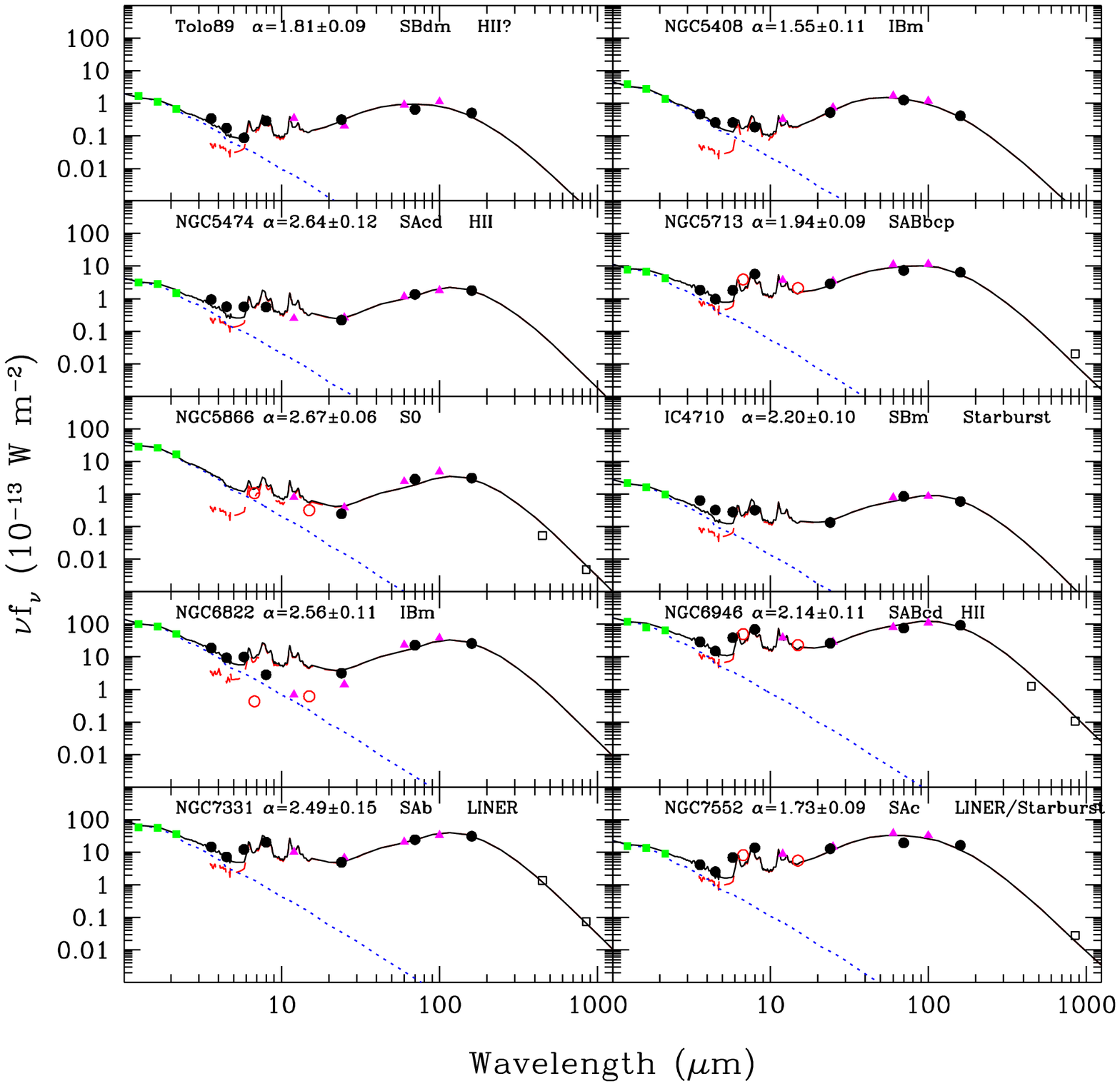}
 \caption{Similar to Figure~\ref{fig:seds1} for 10 more SINGS galaxies.}
 \label{fig:seds7}
\end{figure}

\begin{figure}
 \plotone{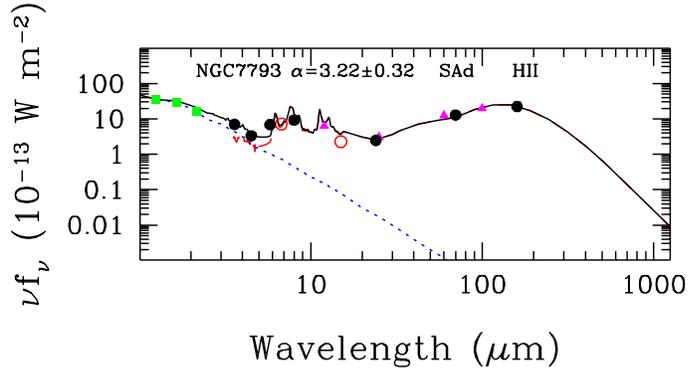}
 \caption{Similar to Figure~\ref{fig:seds1} for 1 more SINGS galaxy.}
 \label{fig:seds8}
\end{figure}

\begin{figure}
 \plotone{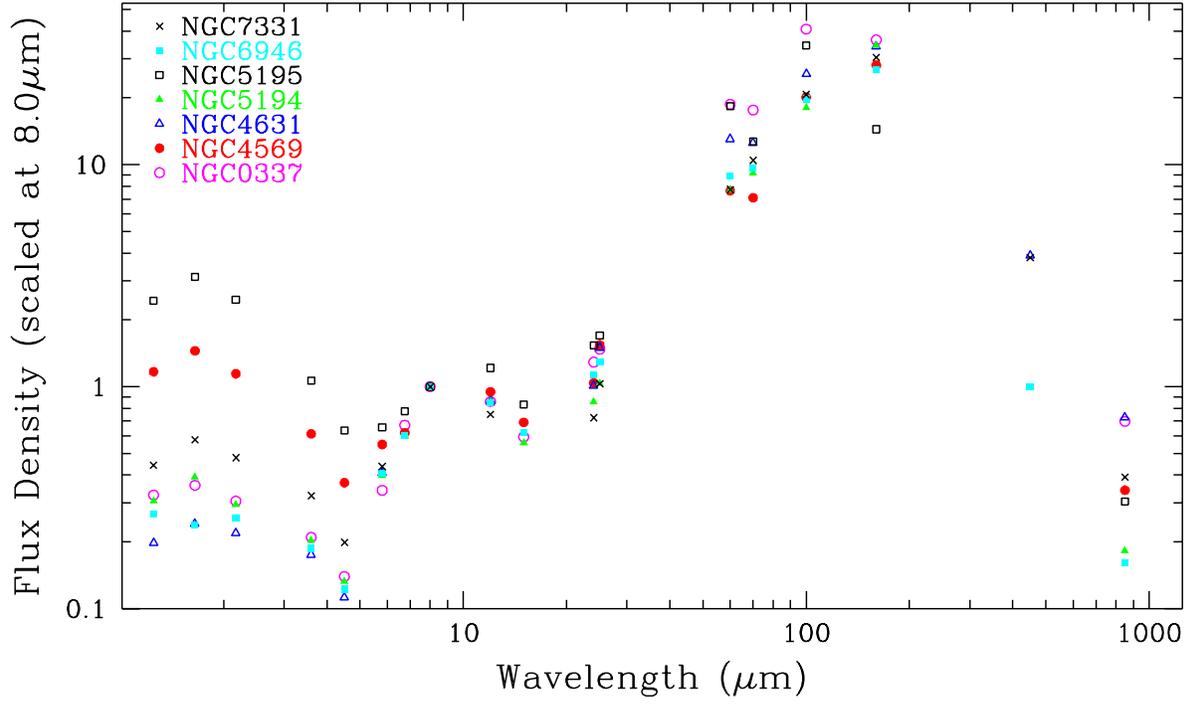}
 \caption{A representative collection of empirical SEDs from the SINGS sample.  Normalizing at 8.0\m\ shows the dramatic variations over infrared wavelengths, especially in the near-infrared.}
 \label{fig:seds_sample}
\end{figure}

\begin{figure}
 \plotone{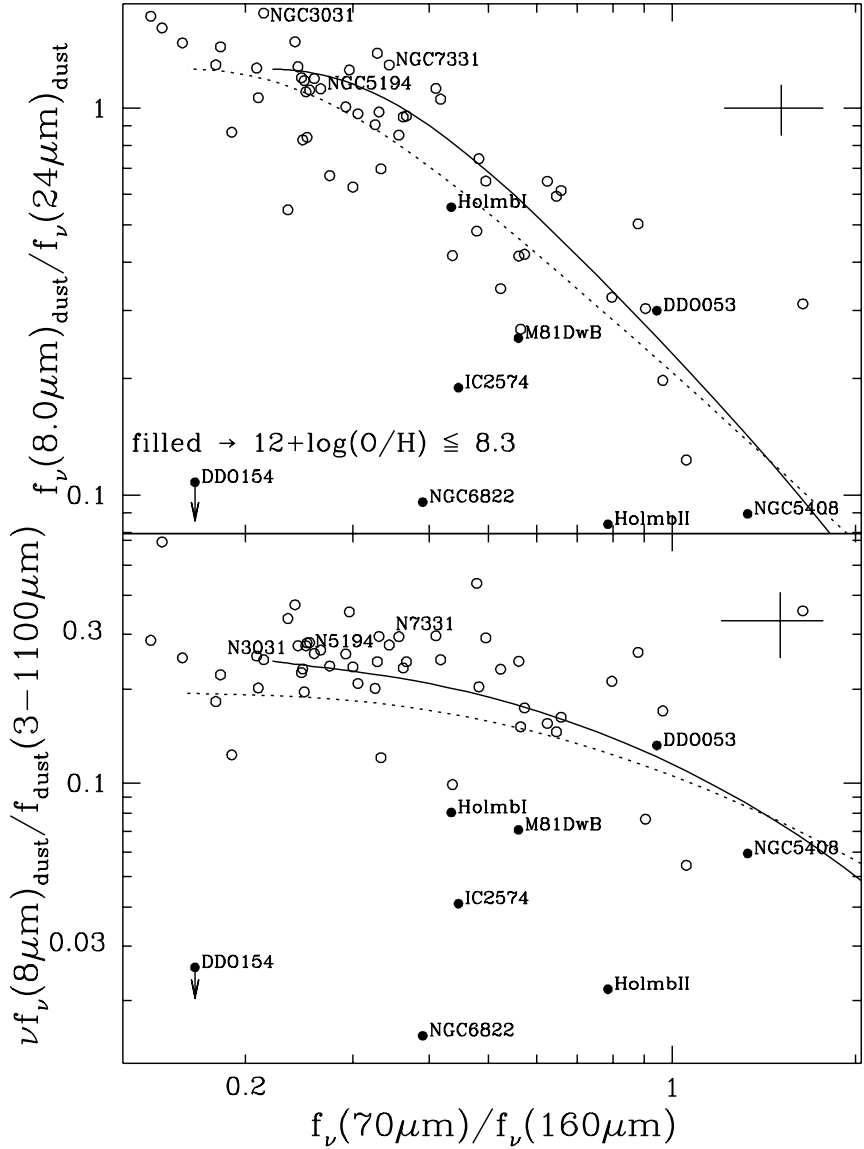}
 \caption{Top: The {\it Spitzer} (dust-only) mid- and far-infrared color-color diagram for globally-integrated SINGS data.  Filled symbols represent low metallicity galaxies.  The solid and dotted lines indicate the dust-only SED models of Dale \& Helou (2002) and Dale et al. (2001), respectively, derived from the average global trends for a sample of normal star-forming galaxies observed by \ISO\ and \IRAS.  Bottom: The 8.0\m\ (dust-only) flux with respect to the total 3-1100\m\ dust emission, as a function of far-infrared color.  Note that only low-metallicity galaxies are significantly below the normal star-forming galaxy dust curves (3$\sigma$ upper limits are provided for DDO~154).  A typical set of error bars are provided in both panels for reference.}
 \label{fig:global_colors}
\end{figure}

\begin{figure}
 \plotone{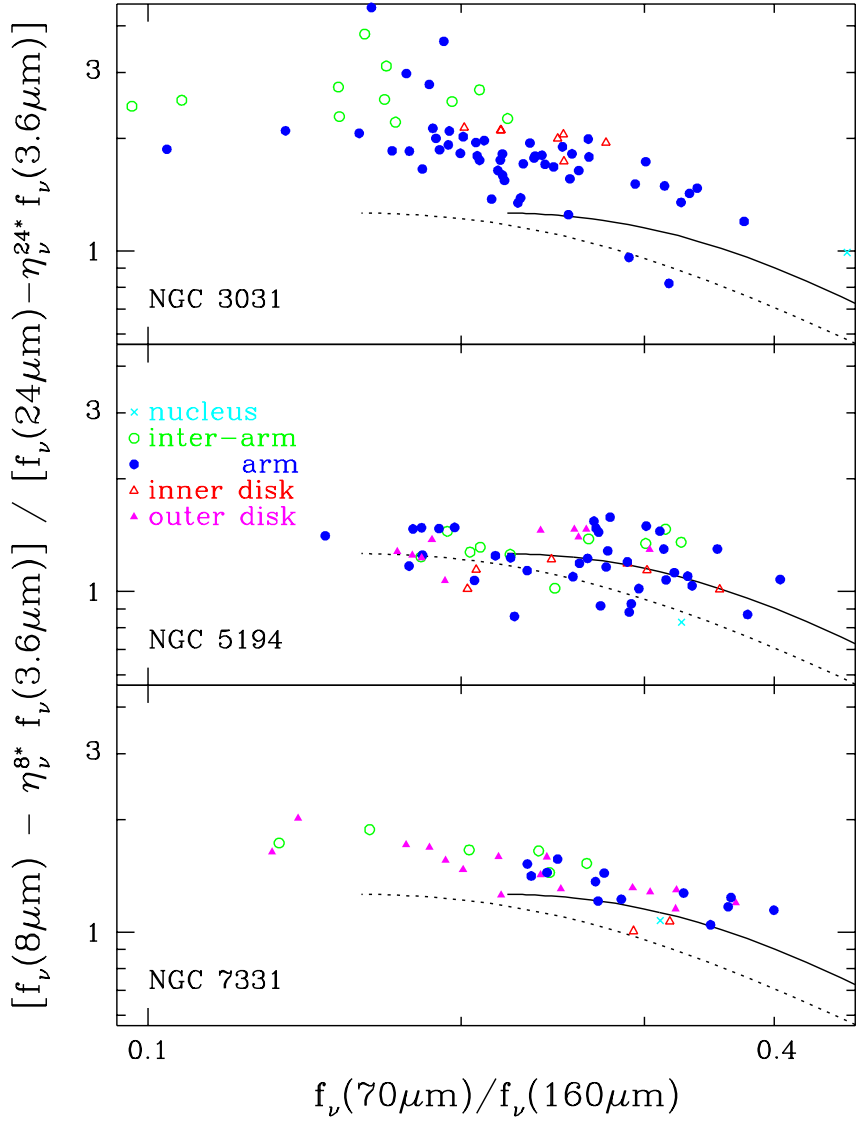}
 \caption{The {\it Spitzer} (dust-only) mid- and far-infrared color-color diagrams for 170$+$ regions within NGC~3031, NGC~5194, and NGC~7331.  The solid and dotted lines indicate the (globally-averaged) SED models of Dale \& Helou (2002) and Dale et al. (2001), respectively.  The local variations within galaxies follow the shape of the global trends (but sometimes at an offset), and the largest variations are seen for the source observed at the highest spatial resolution, NGC~3031.}
 \label{fig:local_colors}
\end{figure}

\begin{figure}
 \plotone{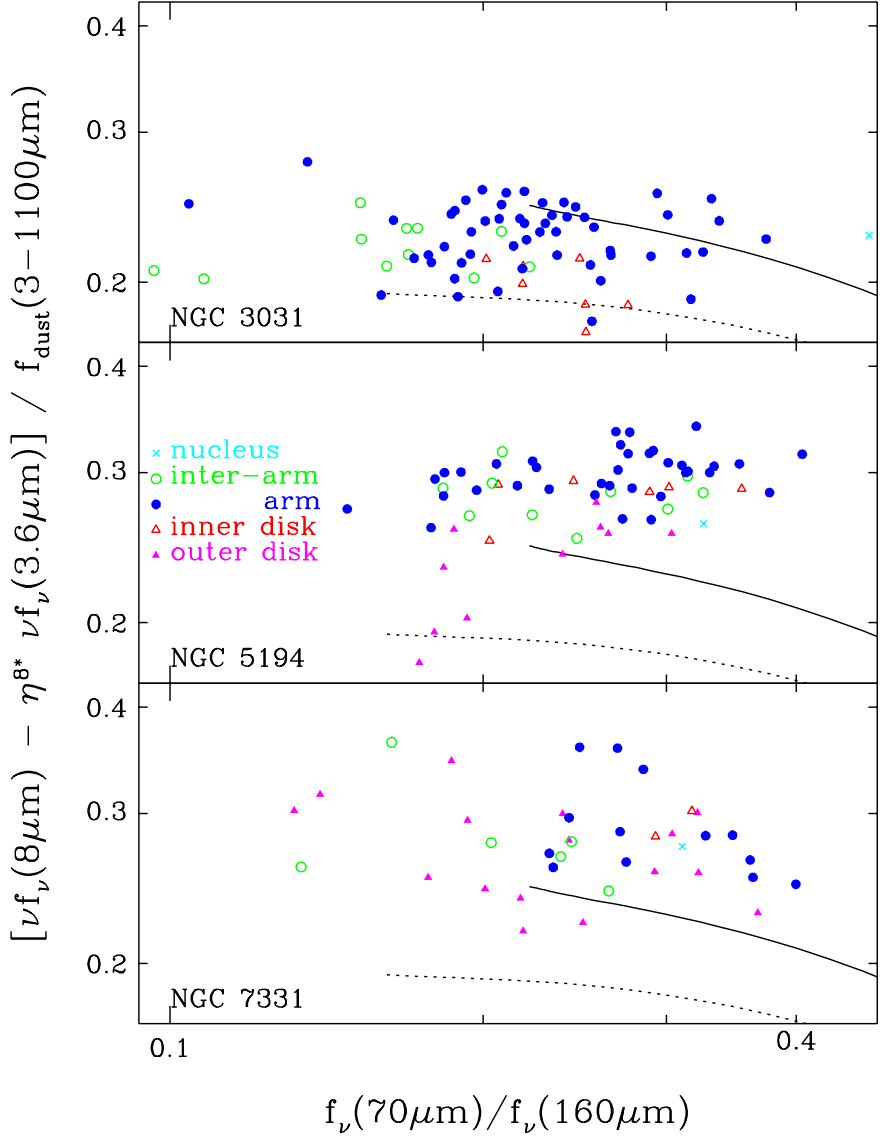}
 \caption{The ratio of dust-only 8.0\m\ emission to the total 3-1100\m\ dust emission as a function of the far-infrared color.  The relative constancy in the ratio is consistent with the interpretation that elevated 24\m\ emission (not reduced PAH emission) drives the \Spitzercolorc\ ratio to low values in regions of higher dust temperature in these three galaxies.}
 \label{fig:local_dust}
\end{figure}

\begin{figure}
 \plotone{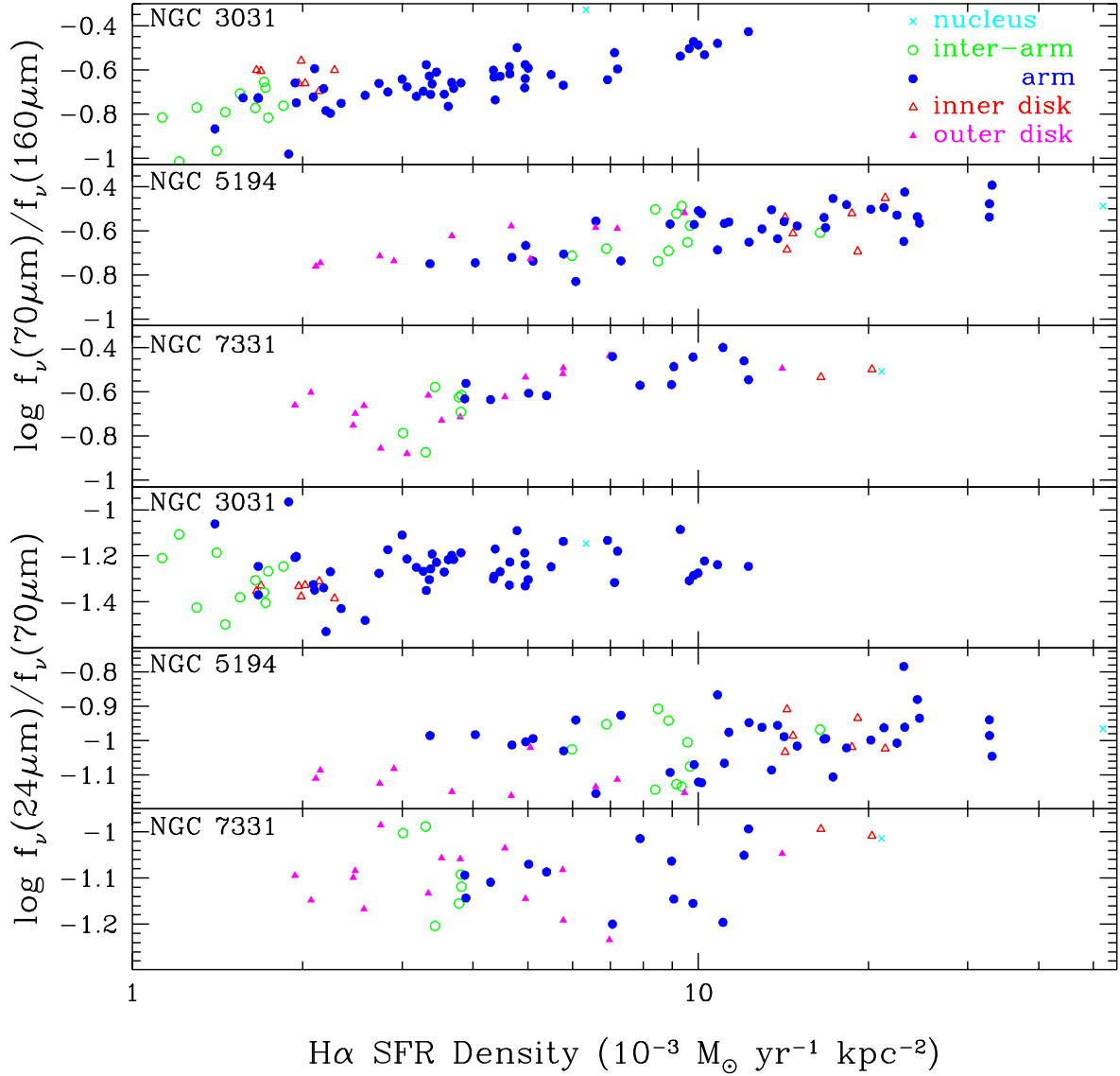}
 \caption{The {\it Spitzer} local far- and mid-infrared colors as a function of the star formation rate surface density.  The far-infrared color correlates with the star formation activity level within these galaxies at the $4-7 \sigma$ level, whereas the mid-infrared color is less correlated at the $1-3 \sigma$ level.}
 \label{fig:sfr}
\end{figure}

\clearpage
\begin{figure}
 \plotone{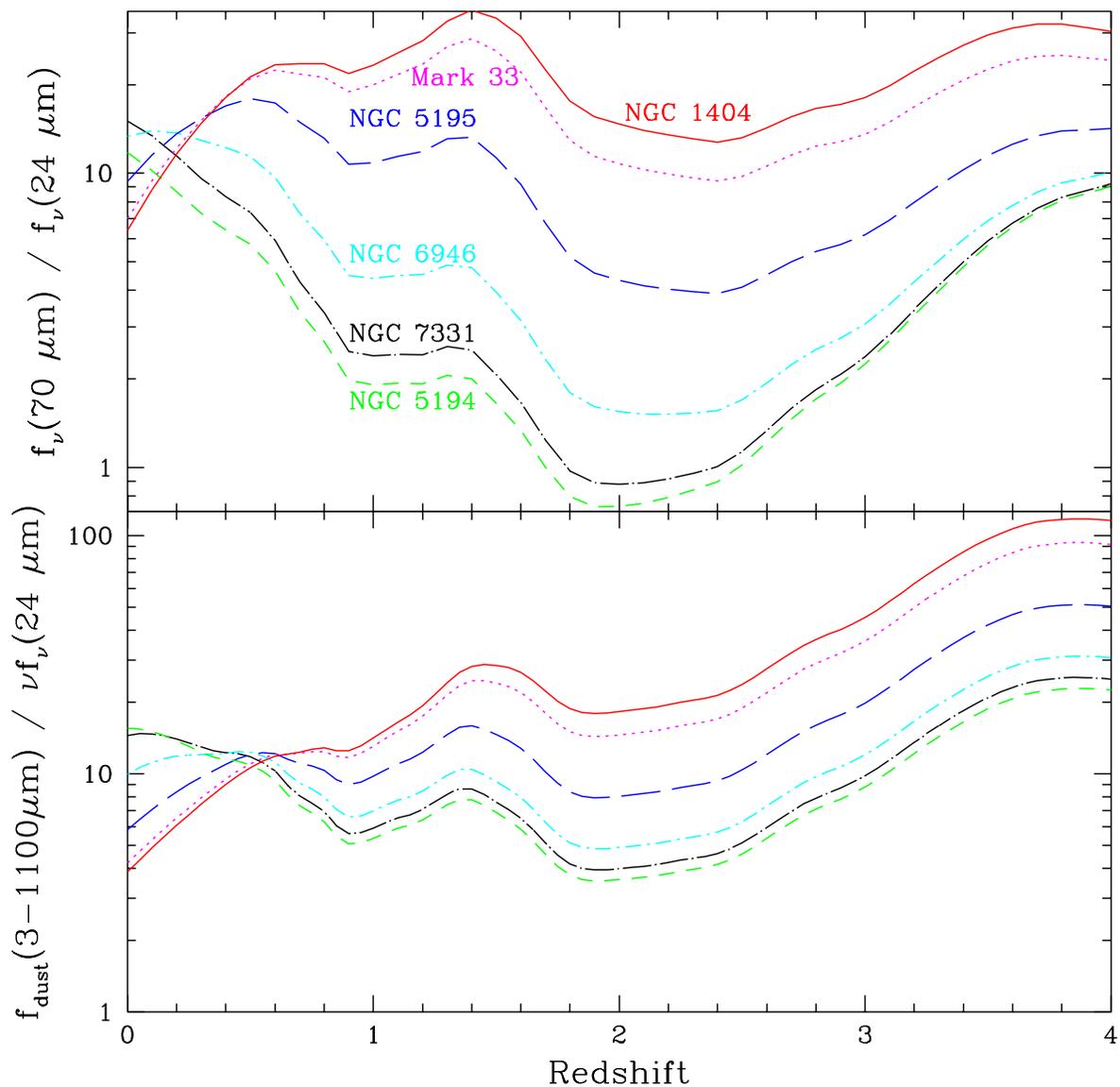}
 \caption{Top: The range of {\it Spitzer} \Spitzercolorf\ colors as a function of redshift, predicted for systems similar to SINGS galaxies.  Bottom: The (dust) bolometric infrared flux, scaled to the observed 24\m\ flux, as a function of redshift.}
 \label{fig:fir_z}
\end{figure}
\end{document}